\documentclass{aa}  
\pdfoutput=1
\usepackage{graphicx}
\usepackage{txfonts}
\usepackage[]{natbib} 
\begin{document}

   \title{Dynamical signatures of a $\Lambda$CDM-halo and the distribution of the baryons in M33}  
 
  \titlerunning{Signatures of a $\Lambda$CDM dark halo in M33} 
  \authorrunning{E. Corbelli et al.} 
 
   \author{Edvige Corbelli,
          \inst{1}
          David Thilker,\inst{2}
          Stefano Zibetti\inst{1}
	  Carlo Giovanardi,\inst{1}
          \and
	  Paolo Salucci\inst{3} 
           }

   \institute{INAF-Osservatorio Astrofisico di Arcetri, Largo E. Fermi, 5,
             50125 Firenze, Italy\\
              \email{edvige,zibetti,giova@arcetri.astro.it}
         \and
             Center for Astrophysical Sciences,The Johns Hopkins University, 
             3400 N.Charles Street, Baltimore, MD 21218, USA\\
             \email{dthilker@pha.jhu.edu}
         \and
             Department of Astrophysics, SISSA, Via Beirut, 2-4, 34014 Trieste, Italy\\
             \email{salucci@sissa.it}
             }

   \date{Received .....; accepted ....}

 \abstract
   {}
   { We determine the mass distribution of stars, gas and dark matter in the nearby galaxy M33
    to test cosmological models of galaxy formation and evolution.
   }
   { We map the neutral atomic gas content of M33 using high resolution VLA and 
    GBT observations of the 21-cm HI line emission. A tilted ring model is fitted to the 
    HI datacube to determine the varying spatial orientation of the extended gaseous disk and its rotation 
    curve. We derive the stellar mass surface density map of M33's optical disk via pixel-SED fitting methods based on
     population synthesis models which allow for positional changes in star
     formation history, and estimate the stellar mass surface density beyond the optical edge from deep 
    images of outer disk fields. Stellar and gas maps are then used in the dynamical analysis of the rotation
    curve to constrain the dark halo
    properties in a more stringent way than previously possible.}
   { The disk of M33 warps from 8 kpc outwards without substantial
     change of its inclination with respect to
     the line of sight, but rather in a manner such that the line of nodes rotates clockwise towards the direction of M31. 
     Rotational velocities rise steeply with radius in the inner disk, reaching 100~km~s$^{-1}$ in 4~kpc, 
     then the rotation curve becomes more perturbed and flatter with velocities as high as  
     120-130~km~s$^{-1}$ out to 2.7~R$_{25}$. The stellar surface density map highlights a
     star-forming disk whose mass is dominiated by stars with a
     varying mass-to-light ratio.  At larger radii a dynamically relevant fraction of 
     the baryons are in gaseous form. A dark matter halo with a Navarro-Frenk-White density profile, as predicted by
     hierarchical clustering and structure formation in a $\Lambda$CDM cosmology, provides the best fits to the rotation
     curve. Dark matter is relevant at all radii in shaping the rotation curve and the most likely dark halo has a concentration  
     C$\simeq$10  and a total mass of 4.3($\pm$ 1.0) 10$^{11}$~M$_\odot$. 
     This imples a baryonic fraction of order 0.02  and the
     evolutionary history of this galaxy should therefore account for loss of a 
     large fraction of its original baryonic content. }
  {}

   \keywords{Galaxies: individual (M\,33) --
             Galaxies: ISM --
         Galaxies: kinematics and dynamics --
         radio lines: galaxies --
         dark matter
            }

   \maketitle

\section{Introduction}

Rotation curves of spiral galaxies are fundamental tools to study the
visible mass distributions in galaxies and to infer the properties of
any associated dark matter halos. Knowledge of the radial
halo density profile from the center to the outskirts of galaxies is
essential for solving crucial issues at the heart of galaxy formation
theories, including the nature of the dark matter itself.  Numerical
simulations of structure formation in the flat Cold Dark Matter
cosmological scenario (hereafter $\Lambda$CDM) predict a well defined
radial density profile for the collisionless particles in virialized
structures, the NFW profile \citep{1996ApJ...462..563N}. 
The two parameters of the
``universal'' NFW density profile are the halo overdensity and the
scale radius, or (in a more useful parameterization) the halo
concentration and its virial mass. For hierarchical structure
formation, small galaxies should show the highest halo concentrations
and massive galaxies the lowest ones \citep{2007MNRAS.378...55M}. 
The relation between these parameters is still under investigation since
numerical simulations improve steadily by considering more stringent
constraints on the cosmological parameters, by using a higher numerical 
resolution and by considering phenomena which can affect the growth or time 
evolution of dark matter halos. 
$\Lambda$CDM dark halo models have been shown to be appropriate for explaning the rotation
curve of some spiral galaxies \citep[e.g. ][]{2013A&A...557A.131M} including that of the nearest spiral, 
M31, which has been traced out to large galactocentric distances by the motion of its satellites 
\citep{2010A&A...511A..89C}. Dwarf galaxies with
extended rotation curves have instead contradicted Cold Dark
Matter scenarios since central regions show constant density cores 
\citep{2005ApJ...634L.145G,2007MNRAS.375..199G}.
Shallow central density distributions  and very low dark matter concentrations
have been supported also by high  
resolution analysis of rotation curves of low surface
brightness galaxies \citep{2001ApJ...552L..23D,2004MNRAS.351..903G}.  
This has cast doubt on the nature of dark matter on one hand, but has also given 
new insights into the possibility that the halo structure might have been altered by  
galaxy evolution \citep[e.g.,][]{2002MNRAS.333..299G}.   
Recent hydrodynamical simulations in $\Lambda CDM$ framework have also shown that
strong outflows in dwarf galaxies inhibit the formation of bulges and
decrease the dark matter density, thus reconciling dwarf galaxies with
$\Lambda CDM$ theoretical predictions \citep{2009MNRAS.398..312G,2010Natur.463..203G}.
Also, shallower density profiles than true ones might be recovered in the presence of 
small bulges or bars \citep{2004ApJ...617.1059R}.
M33 provides useful tests for the $\Lambda CDM$ cosmological scenario  since 
it is higher in mass than a dwarf, but it hosts no
bulge nor prominent bars \citep{2007ApJ...669..315C}. The absence of prominent
baryonic mass concentrations  alleviates the uncertainties in the inner circular velocities and 
makes it unlikely that the baryonic collapse and disk assembly alter the distribution of dark matter 
in the halo. 
 
M33 is a low-luminosity spiral galaxy, the most common type of spiral in the universe \citep{1999ApJ...521...50M} 
and the third most luminous member of the Local Group. It yields a unique opportunity to study the
distribution of dark matter because it is richer in gas, and more dark matter dominated than our
brighter neighborhood M31. Due to its proximity,
the rotation curve can be traced with unprecedented spatial 
resolution. M33 is in fact one of the very few  objects for which it has been possible to combine 
a high quality and high resolution  extended rotation curve \citep{2000MNRAS.311..441C} with a well determined
distance \citep[assumed in this paper to be D = 840~kpc ][]{1991ApJ...372..455F,2013ApJ...773...69G},
and with an overwhelming quantity of data  across the electromagnetic spectrum. 
Previous analysis have shown that the HI and CO velocity fields are very 
regular and cannot be explained by Newtonian dynamic without including a massive dark 
matter halo \citep{2003MNRAS.342..199C}. On the other hand the dark matter density distribution 
inside the halo is still debated due to the unconstrained value of the mass-to-light ratio
of the stellar disk. These uncertainties leave open the possibility to fit the rotation curve
using non-Newtonian dynamic in the absence of dark matter \citep{2007MNRAS.374.1051C}.  

The extent of the HI disk of M33 has been determined through several deep 21-cm observations in the past
\citep[e.g.][]{1978IAUS...77..197H,1989AJ.....97..390C,1997ApJ...479..244C,2009ApJ...703.1486P}. 
Beyond the HI disk,  
\citet{2008A&A...487..161G} found some discrete HI clouds but there is no evidence of a ubiquitous HI
distribution with N$_{HI}< 1-2 \times 10^{19}$~cm$^{-2}$. This is likely due to a sharp HI-HII transition as the
HI column density becomes mostly ionized by the UV extragalactic background radiation \citep{1993ApJ...419..104C}.
Previous observations of the outer disk have been carried out at low spatial resolution, mostly with single
dish radio telescopes  (FWHM$>$3~arcmin) which needed sidelobe corrections. 
The new VLA+GBT HI survey of M33, presented in this paper, has an unprecedented sensitivity and spatial
resolution. It is one of the most detailed databases
ever made of a full galaxy disk, including our own. Our first aim is to use the combined VLA+GBT 
HI survey  to better constrain the amplitude and orientation of the warp  and the rotation curve.  The final 
goal will be to  constrain the baryonic content of M33 disk and the  distribution of dark 
matter in its halo through the dynamical analysis of the rotation curve. 
 
There are also crucial, but now questionable, assumptions that were made in the prior analyses which leads us to further investigate the
baryonic and dark matter content of M33 once again.  Firstly,  a constant mass-to-light ratio has been used throughout 
the disk and  determined via a dynamical analysis of the rotation
curve. Futhermore, the light distribution
has also been fitted using one value of the exponential scalelength in the K-band.
This might not be consistent with the negative radial metallicity 
gradient which supports an inside-out formation 
scenario  \citep{2007A&A...470..843M,2010A&A...512A..63M,2011ApJ...730..129B}.  
\citet{2010A&A...521A..82P} investigated  the radial variations of the stellar mass-to-light
ratio using chemo-photometric models and galaxy colors. They indeed found a radially
decreasing value which affects the mass modeling. 
Evidence of a radially varying mass-to-light ratio has been found also through 
spectral synthesis techniques in large samples of disk dominated galaxies
\citep{ZCR09,2014A&A...562A..47G}. In order to overcome this
simplified constant M/L ratio model, and given the proximity of M33, in this
paper we shall build a surface density map for the bright star forming disk of M33.
Deep optical surveys of the outskirts of M33 have shown that the outer disk is not devoid 
of stars and that several episodes of star formation have occurred \citep[e.g.]
[and references therein]{2011ApJ...728L..23D,2011A&A...533A..91G, 2011A&A...534A..96S}. 
It is therefore important to account for the stellar density and radial scalelength 
in the outer disk of M33  before investigating the dark matter contribution to the outer rotation curve.

In Section 2, we describe the data used to retrieve the velocities
and gas surface densities along the line of sight. The details of the reduction 
and analysis of the optical images used to build the stellar surface density map
are given in Section 3. In Section 4 we  
infer the disk orientation and the rotation curve. Some details of the warp and disk thickness
modeling procedures are given in the Appendix. In Section 5 we derive
the surface density of the baryons. The rotation curve dynamical mass 
models and the dark halo parameters are  discussed in Section 6, also in framework of 
the $\Lambda$CDM scenario. 
Section 7 summarizes the main results on the  baryonic and dark matter distribution
in this nearby spiral.

\section{The gaseous disk}

The HI data products utilized for our study originate from a coordinated observing campaign 
using the Very Large Array (VLA) and the Green Bank Telescope (GBT). Multi-configuration VLA 
imaging was obtained for programs AT206 (B, CS) and AT268 (D-array), conducted in 1997-1998 
and 2001, respectively.    During the AT206 (B, CS-array) observations, we employed a six-point 
mosaicing strategy, with the mosaic spacing and grid orientation designed to provide relatively 
uniform sensitivity over the high-brightness portion of the M33 disk (r $< $ 9.5~kpc i.e. 39~arcmin).  
The D configuration observations (AT268) were obtained with the VLA antennas continuously scanning 
over a larger, rectangular region centered on M33.  A grid of 99 discrete phase tracking centers was 
utilized for this on-the-fly (OTF) observing mode.  The aggregate visibility data from all three 
(B, CS, and D) VLA configurations is sensitive to HI structures at angular scales of $\sim$ 4--900 arcsec.  
Early results based on the B and CS configuration data were presented in \citet{2002ASPC..276..370T},
while an independent reduction of all three VLA configurations has been presented by \citet{2010A&A...522A...3G}.  
The datacubes analyzed herein benefit from combination of the VLA observations with GBT total power data, 
and hence should be considered superior to all datasets published earlier, with the sole exception of 
\citet{2012ApJ...749...87B} for which the dataset is identical.

Our single dish GBT observations were obtained during 2002 October, using the same setup described by 
\citet{2004ApJ...601L..39T} for M31 observations. The telescope was scanned over a 5$\times$5 deg region centered on M33 
while HI spectra were measured at a velocity resolution of 1.25 km~s$^{-1}$. At M33’s distance of 840~kpc, 
the GBT beam (9'.1 FWHM) projects to 2.2~kpc, and our total power survey covers a  74$\times$74 kpc$^{-2}$ region. 
At a resolution of 3~kpc and 18~km~s$^{-1}$, the achieved rms flux sensitivity was 7.2 mJy/beam, corresponding 
to an HI column density of 2.5$\times 10^{17}$~cm$^{-2}$ in a single channel averaged over the beam.   Joint 
deconvolution of all of the VLA interferometric data was carried out after inclusion of the appropriately 
scaled GBT total power images following the method employed for the reduction of the M31 data described 
in detail by \citet{2009ApJ...695..937B}. 

The angular resolution of the HI products used for our study was substantially less than the limit of our 
VLA+GBT dataset.  In practice, we used datacubes having 20~arcsec (81~pc) and 130~arcsec (0.53~kpc) restored beam size.  
Masking of these cubes was completed in order to limit our analysis to regions with sufficient 
signal-to-noise and mitigate contributions from foreground Milky Way emission.  
With the support from masked regions outside of M33, we interpolated Galactic emission structures 
across M33 and subtracted this component to account crudely for its presence.

From the masked 20" and 130" datacubes we produced moment-0, moment-1, peak intensity, and velocity 
at peak intensity images.  These image  products were generated in the usual manner with  Miriad's {\tt moment} task.  
The peak intensity and velocity at peak intensity images are derived from a  three-point quadratic fit. Finally,  
Miriad's {\tt imbin} task was used to generate re-binned cubes with only one pixel per resolution element, such 
that adjacent pixels were uncorrelated. These binned cubes were the products used to constrain the galaxy model.

\begin{figure*} 
\centerline{
\includegraphics[width=8cm]{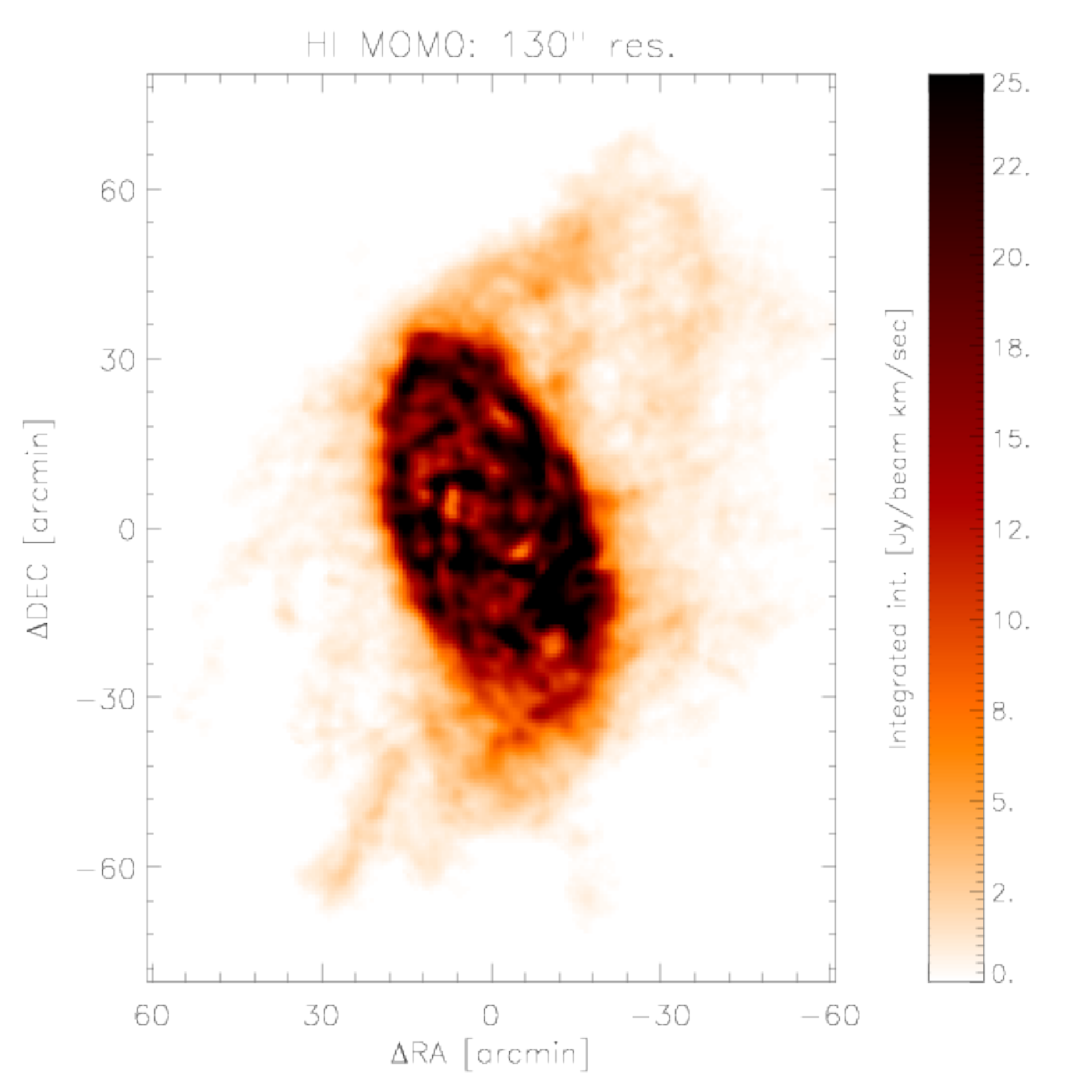}
\includegraphics[width=8cm]{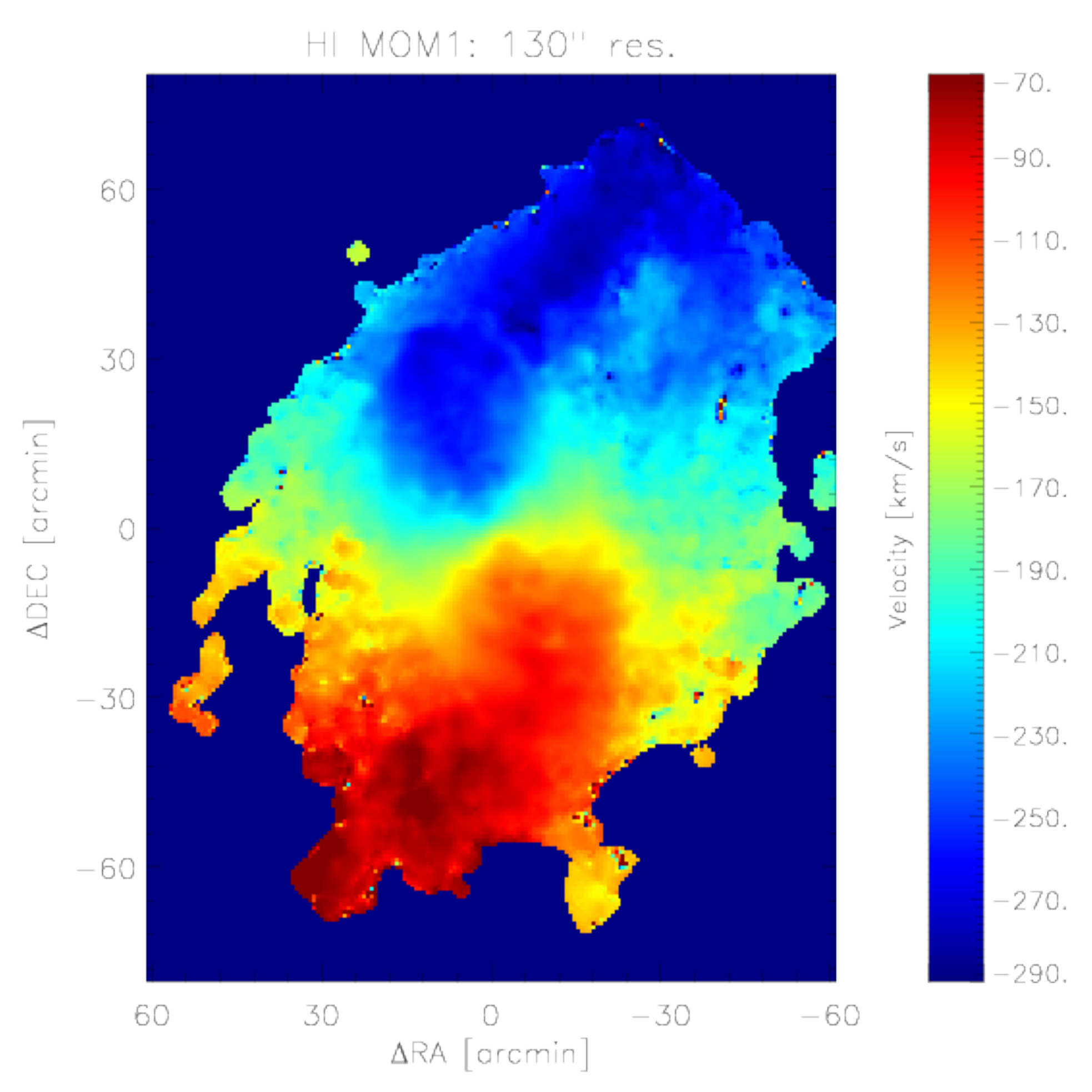}}
\centerline{
\includegraphics[width=8cm]{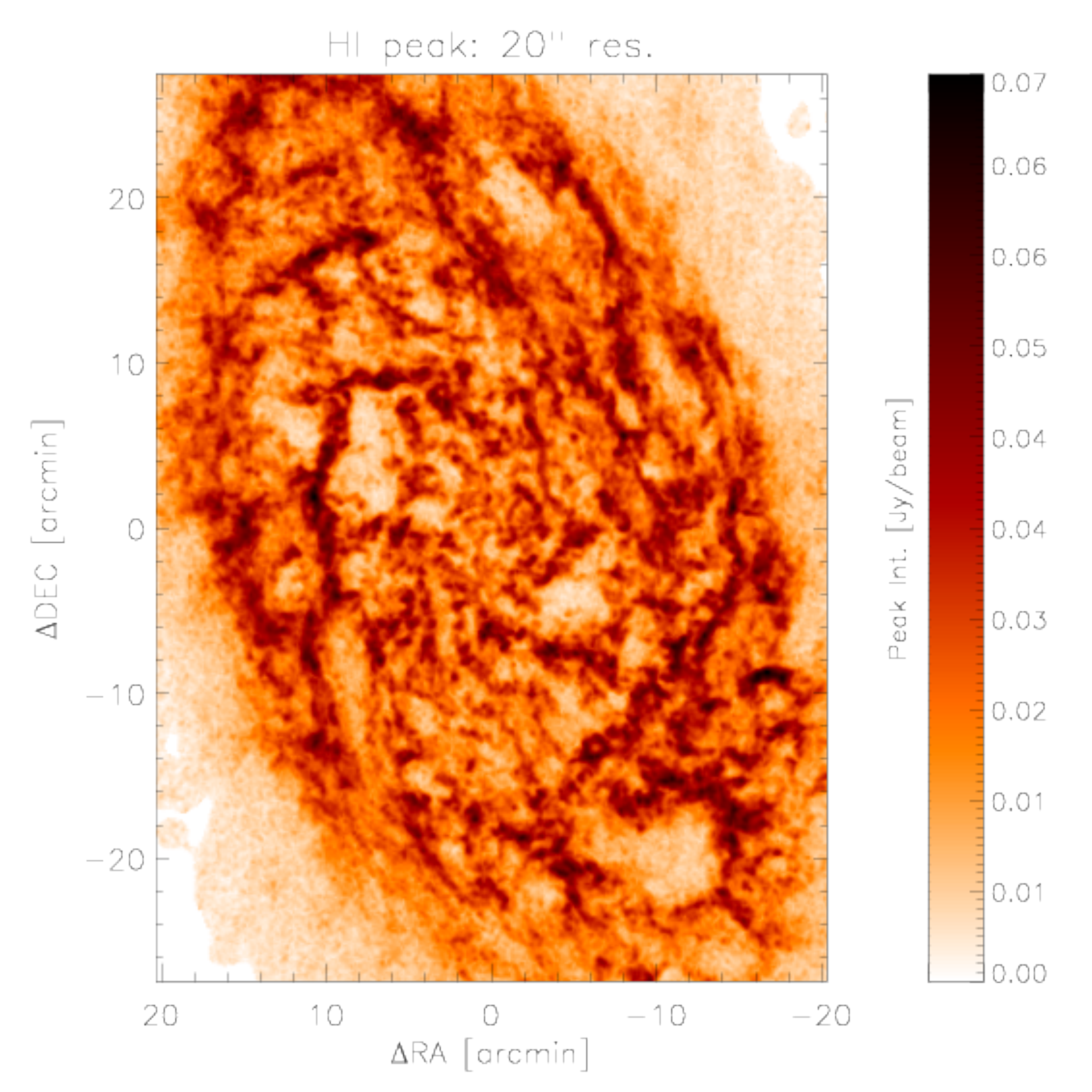} 
\includegraphics[width=8cm]{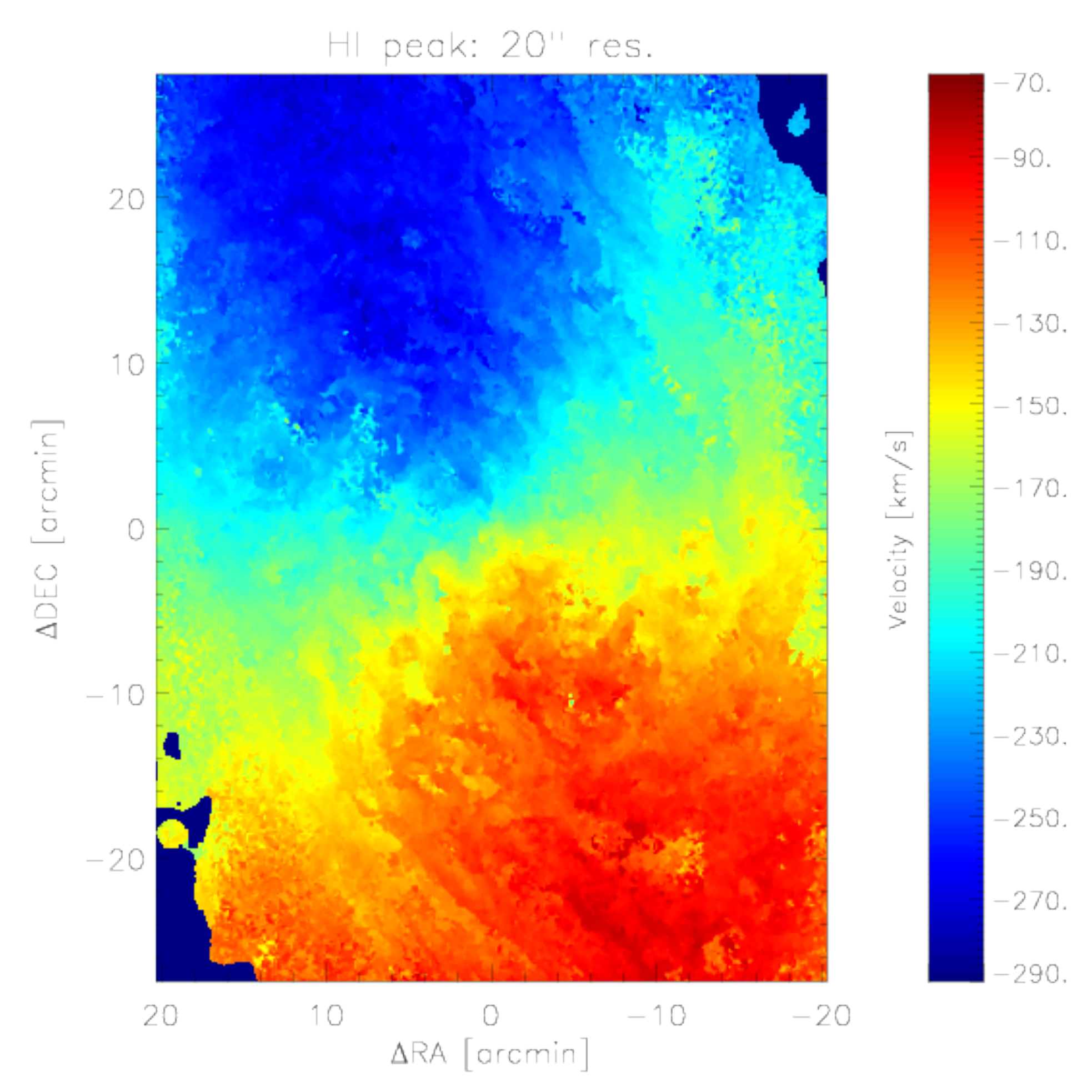}}
\caption{In the upper panels we show the integrated  21-cm line intensity and the intensity weighted mean
velocity over the whole galaxy using the data at 130~arcsec resolution. In the lower panels we show the peak
intensity and the velocity at peak maps using the VLA+GBT data at a resolution of 20~arcsec.
}
\label{vladata} 
\end{figure*}

\subsection{The molecular gas map}

Rotational velocities in the innermost regions are complemented by the independent CO J=1-0 line dataset described 
by \citet{2003MNRAS.342..199C} at a spatial resolution of 45~arcsec (FCRAO survey). For the H$_2$ surface mass density 
radial scalelength we use the mean value between that determined by the FCRAO survey and that determined via the
IRAM-30m CO J=2-1 line map of \citet{2010A&A...522A...3G} (1.9 and 2.5~kpc respectively) i.e. we  use the following 
expression to fit the molecular gas surface density:

\begin{equation}
\Sigma_{H_2}=10 \times e^{-R/2.2} \qquad M_\odot~pc^{-2}
\end{equation}

\noindent
where $R$ is in kpc. The total gas mass in molecular form according to the above expression 
is  3.0$\times$10$^8$~M$_\odot$, if integrated through the whole HI extent. This is in agreement with the molecular
mass  inferred by \citet{2003MNRAS.342..199C}, \citet{2010A&A...522A...3G} and more recently by \citet{2014arXiv1405.5884D}.  

The atomic and molecular hydrogen gas surface densities have been multiplied by 1.33 to account for the helium
mass when deriving the total gas mass surface density.

\section{The stellar mass surface density}

In this section we describe how the stellar mass surface density maps
are derived: in \S\ref{mstar_method} we present the method, then in
\S\ref{mstar_SDSS} and \ref{mstar_LGS} we describe the data and their
processing and finally in \S\ref{mstar_maps} we present the
results. The extension of the stellar mass distribution to distances
larger than $\sim 5$~kpc is described in \S\ref{mstar_extern}.

\subsection{Method: from light to stellar mass}\label{mstar_method}

Maps of stellar mass surface densities can be derived from multiband
optical imaging, as shown in the seminal work of \citet[][ZCR09
hereafter]{ZCR09}, thanks to the relatively tight correlation existing
between colors and the apparent stellar mass-to-light ratio
\citep[$M/L$, e.g.][]{bell_dejong01}. Here we implement an extension
of the ZCR09 method, based on a fully bayesian approach, similar to the one adopted and validated by a number of
works in the literature of the last decade 
\citep[e.g.][]{2003MNRAS.341...33K,2005ApJ...619L..39S,2005MNRAS.362...41G,2008MNRAS.388.1595D,2009ApJS..185..253G}. 
Our imaging dataset is composed of mosaic maps in the $B$,
$V$ and $I$ bands from the Local Group Survey
\citep{2006AJ....131.2478M} and $g$ and $i$ bands from the Sloan
Digital Sky Survey \citep[SDSS,]{SDSS}. After processing the images as
described in the following subsections, we perform a \emph{pixel by
  pixel} stellar population analysis by comparing the measured
magnitudes in the five bands with the corresponding magnitudes derived
from a comprehensive model library of 150\,000 synthetic spectral
energy distributions (SED). Each model SED is computed with a stellar
population synthesis (SPS) technique based on the most recent revision
of the \citet[][BC03, Charlot \& Bruzual in prep.]{BC03} simple
stellar populations (SSPs), which use the ``Padova 1994'' stellar
evolutionary library (see BC03 for details) and \citet{2003ApJ...586L.133C}
stellar initial mass function (IMF). The SSPs are linearly combined
according to a variety of star formation histories (SFH),
parameterized and distributed as in \citet[][see their Sec. 2.3 for details]{2005MNRAS.362...41G}, where random
(in terms of time of occurrence, intensity and duration) bursts of
star formation are superimposed onto continuous exponentially
declining SFHs, with random initial time $t_\mathrm{form}$ and
e-folding time scales $1/\gamma$. The metallicity of the stellar
populations is fixed in each model along the SFH and is randomly
chosen between 0.02 and 2.5 $Z_\odot$ from a uniform linear
variate. Dust attenuation is also implemented following the simplified
2-component parameterization \`a la \cite{charlot_fall} for the diffuse ISM and the stellar birth clouds: we adopt the same
prescriptions as in \citet[][see their sec. 2.1 for details]{2008MNRAS.388.1595D}.  Each model is originally normalized so
that the total mass in living stars and stellar remnants (white
dwarfs, neutron stars and black holes) at the end of the SFH is 1
$M_\odot$. At each pixel, for each model $i$ we define the chi-square as follows:
\begin{equation}
\chi^2_i=\sum_{j=1}^{Nbands}{\frac{\left(m_{\mathrm{obs},j}-(m_{\mathrm{model},i,j}+ZP_i)\right)^2}{\sigma_j^2}}
\end{equation}
where $j$ runs over all bands used for the mass estimation, $m_{\mathrm{obs},j}$ is the observed 
magnitude of the pixel in the $j$-th band and $\sigma_j$ the corresponding error, 
$m_{\mathrm{model},i,j}$ is the magnitude of the model $i$ in the $j$-th band plus distance modulus, 
and $ZP_i$ is the zeropoint offset to apply to the model to match the observations.
We then compute the $ZP_i$ that minimizes $\chi^2_i$.
Since the models are originally normalized to 1 $\mathrm{M_\odot}$ in stars, the best-fitting stellar 
mass $M_i^*$ associated to the model $i$ is simply given by:
\begin{equation}
M_i^*=10^{(-0.4~ZP_i)}~\mathrm{M_\odot}
\end{equation}
To each model $i$ we further associate a likelihood 
\begin{equation}
\mathcal{L}_i\propto \exp(-\chi_{i,\mathrm{min}}^2/2)
\end{equation}
We can now build the unity-normalized probability distribution function (PDF) for the stellar mass 
at the given pixel: we consider the median-likelihood value (i.e. the value of stellar mass corresponding to the 
cumulative normalized PDF value of 0.5) as our fiducial estimate. The mass range between the 16-th and 84-th 
percentile of the PDF divided by 2 provides the error on the estimate. 
It is worth noting that the so-computed errors account for both photometric uncertainties and, importantly, 
also for intrinsic degeneracies between model colors and M/L.

The adoption of this bayesian median-likelihood estimator,
instead of the simple conversion of light to mass via the M/L derived
from a relation with one or two colors as done in ZCR09, allows to take
full advantage of our multi-band dataset and reduce the
impact of systematic effects in the individual bands (see next
subsection).

We note that the BC03 revised models used here produce significantly
higher stellar masses than those adopted in ZCR09, by some
0.1 dex. The main difference concerns the implementation of the TP-AGB
evolutionary phase, which in the so-called ``CB07'' version predicts
much higher luminosities than in the original BC03 models at fixed
stellar mass. Recent works
\citep[e.g.][]{kriek+10,conroy_gunn,zibetti+13}, however, have shown
that the BC03 models are indeed closer to real galaxies than the CB07
ones and this motivates our choice.

\subsection{The SDSS data}\label{mstar_SDSS}

The Sloan Digital Sky Survey \citep[SDSS,]{SDSS} has covered M33 in
several imaging scans, which must be properly background-subtracted
and combined to obtain a suitable, science-ready map. From the SDSS we use only the
$g$ and the $i$ band, which deliver the best signal-to-noise ratio out
to the optical radius. We discard the $r$ band because of its
contamination by H$\alpha$ line emission (see discussion in ZCR09). We
reconstruct the separate scan stripes that cover M33 by stitching
together the individual fields that we retrieved from the DR8
\citep{SDSS_DR8} data archive server. The SDSS pipeline automatically
subtracts the background by fitting low order polynomials along each
column in the scan direction. The process, however, is optimized to
work with galaxies sizes of few arcminutes at most. In the case of
M33, which is more than 1 degree across, the standard background
subtraction algorithm fails by subtracting substantial fractions of
the galaxy light. Therefore, we have added back the SDSS background
and obtained the original scans, where only the flat-field correction
has been applied. 

For each column along the scan direction we determine the
  background as the best-fitting 5th degree Legendre polynomial, after
  masking out the elliptical region that roughly corresponds to the
  optical extent of M33
  ($80~\mathrm{arcmin}\times60~\mathrm{arcmin}$). This is achieved
  using the \texttt{background} task in the \texttt{twodspec} IRAF
  package, with a 2-pass sigma-clipping rejection at 3 $\sigma$. This
  procedure leaves large scale residuals in the backround regions,
  which we estimate at the level of $\approx 26.5$ mag arcsec$^{-2}$
  in $g$ and $\approx 25.5-26$ mag arcsec$^{-2}$ in $i$ (r.m.s.).
  They are visible in the final mosaic maps as structure along the scan
  direction. We also checked that the extrapolation of the fit over
  the masked central area does not introduce spurious effects.  In
  fact, the amplitude of the fitted polynomial along the scan
  direction never exceeds values corresponding to $\approx 24.5$ mag
  arcsec$^{-2}$ in the central regions, thus impacting their surface
  brightness by a few percent at most.

The scans are then astrometrically registered and combined in a
wide-field mosaic using SWARP \citep{TERAPIX}. Bright Milky Way stars
must be removed before surface photometry can be obtained. Stars are
identified using SExtractor \citep{sextractor} on an unsharp-masked
version of the mosaic.  Bright young stellar association and compact
star forming regions are manually removed, as our intent is to model
the underlying (more massive) stellar disk, and such bright young
features could bias the inferred M/L ratio downward.  The original mosaic is then
interpolated over the detected stars with a constant background value
averaged around each star, using the IRAF task \texttt{imedit}.  The
size of the apertures for interpolation are optimized iteratively in
order to minimize the residuals, and the brightest stars are edited
manually for best results.

The edited SDSS mosaic is then smoothed with a gaussian of $\approx 100$~pc
(25~arcseconds) FWHM in order to remove the stochastic fluctuations
due to bright stars occupying the disk of M33, and finally resampled on a
pixel scale of 8.3 arcsec/pix (i.e. $\approx 3$~pixels per resolution
element). The $g$- and $i$-band surface brightness maps are finally
corrected for the foreground MW extinction, by 0.138 and 0.071 mag
respectively. Thanks to the massive smoothing and resampling, photon
noise is largely negligible. The photometric errors are dominated by
systematics in the photometric zero point and in the background
subtraction. We conservatively estimate these sources of error as 0.05
mag for the zero-point  (including also uncertainties related to
  the unknown variations of foreground Galactic extinction)  and as
30\% of the surface brightness for the background level uncertainty at
$\mu_g=25$ and $\mu_i=24.5~\rm{mag arcsec}^{-2}$   (which takes
  into account the aforementioned fluctuations in the residual
  background), respectively. The two contributions then are added in
quadrature pixel by pixel. The zero-point uncertainty
dominates in the inner and brighter parts of the galaxy, while the
background uncertainty dominates at low surface brightness.

\begin{figure*} 
\centerline{
\includegraphics[width=8cm]{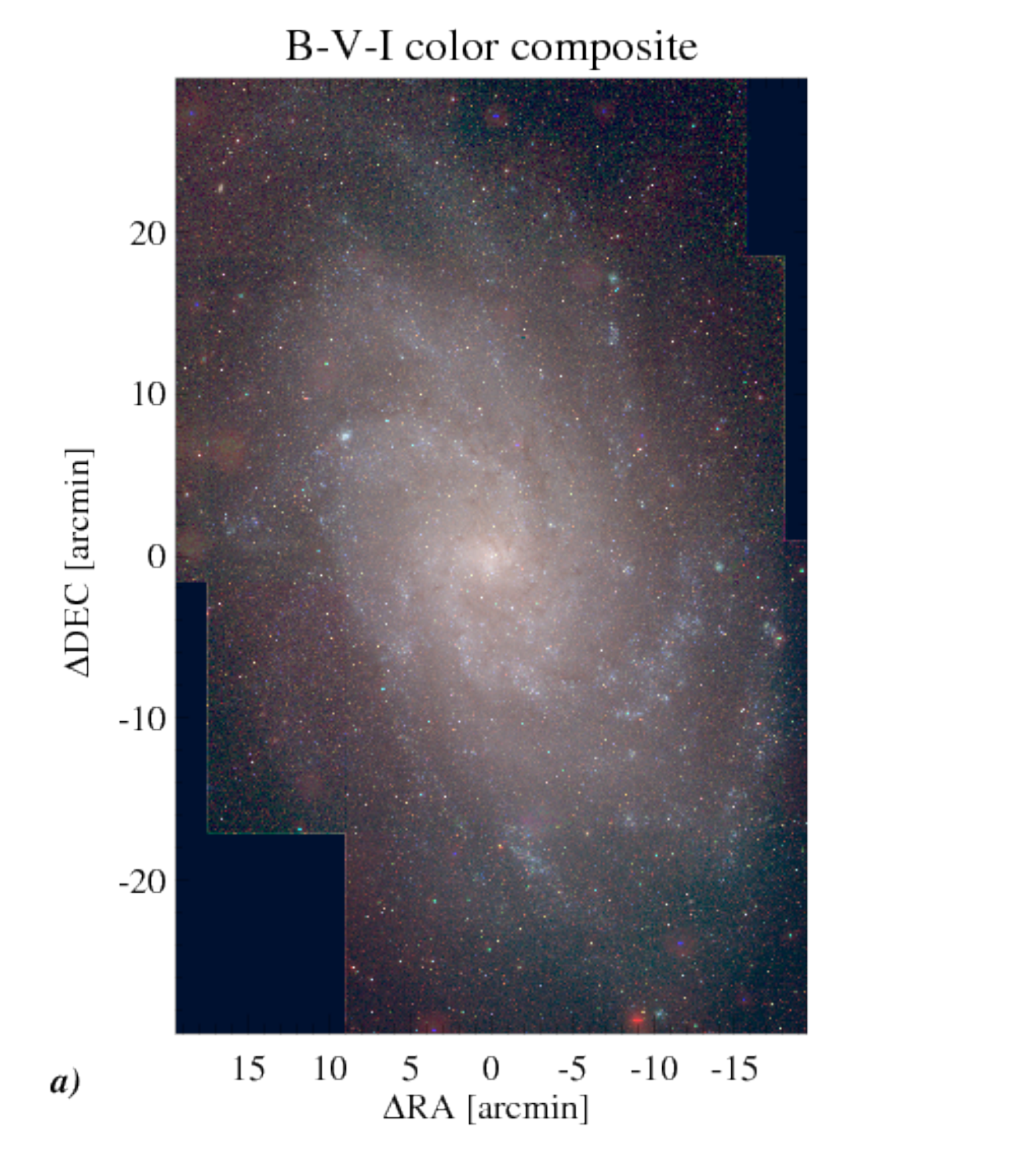}
\includegraphics[width=8cm]{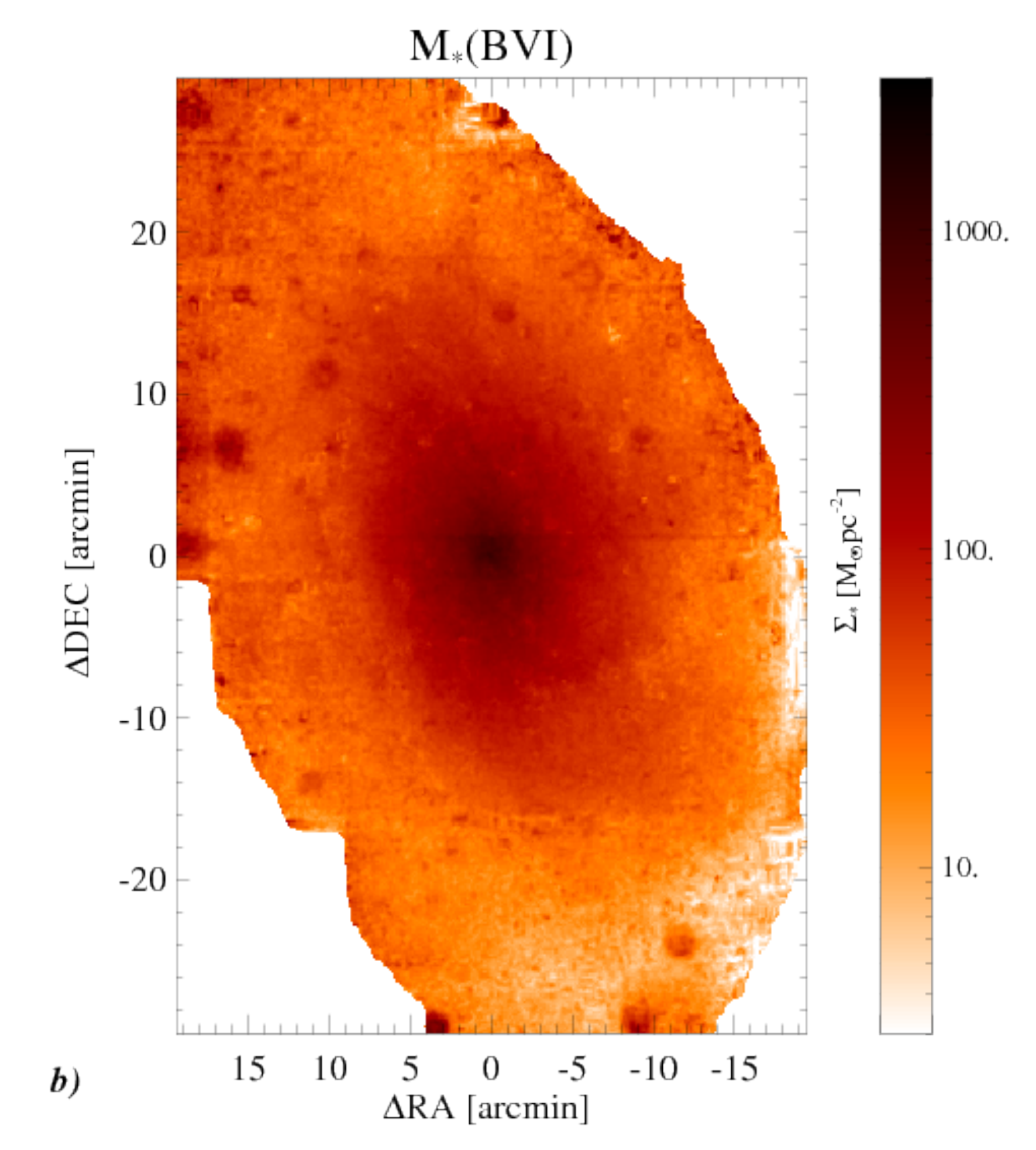}}
\centerline{
\includegraphics[width=8cm]{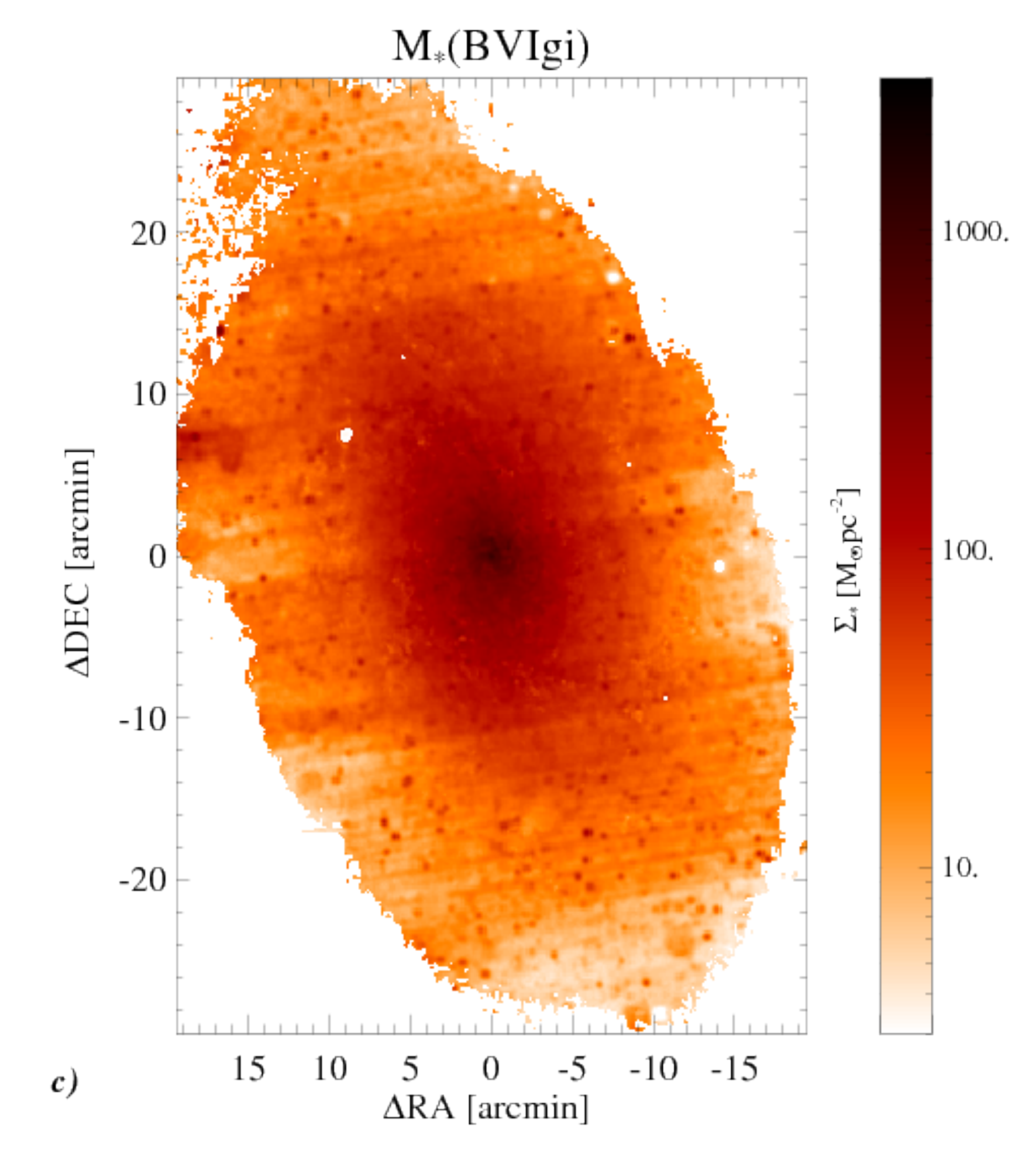} 
\includegraphics[width=8cm]{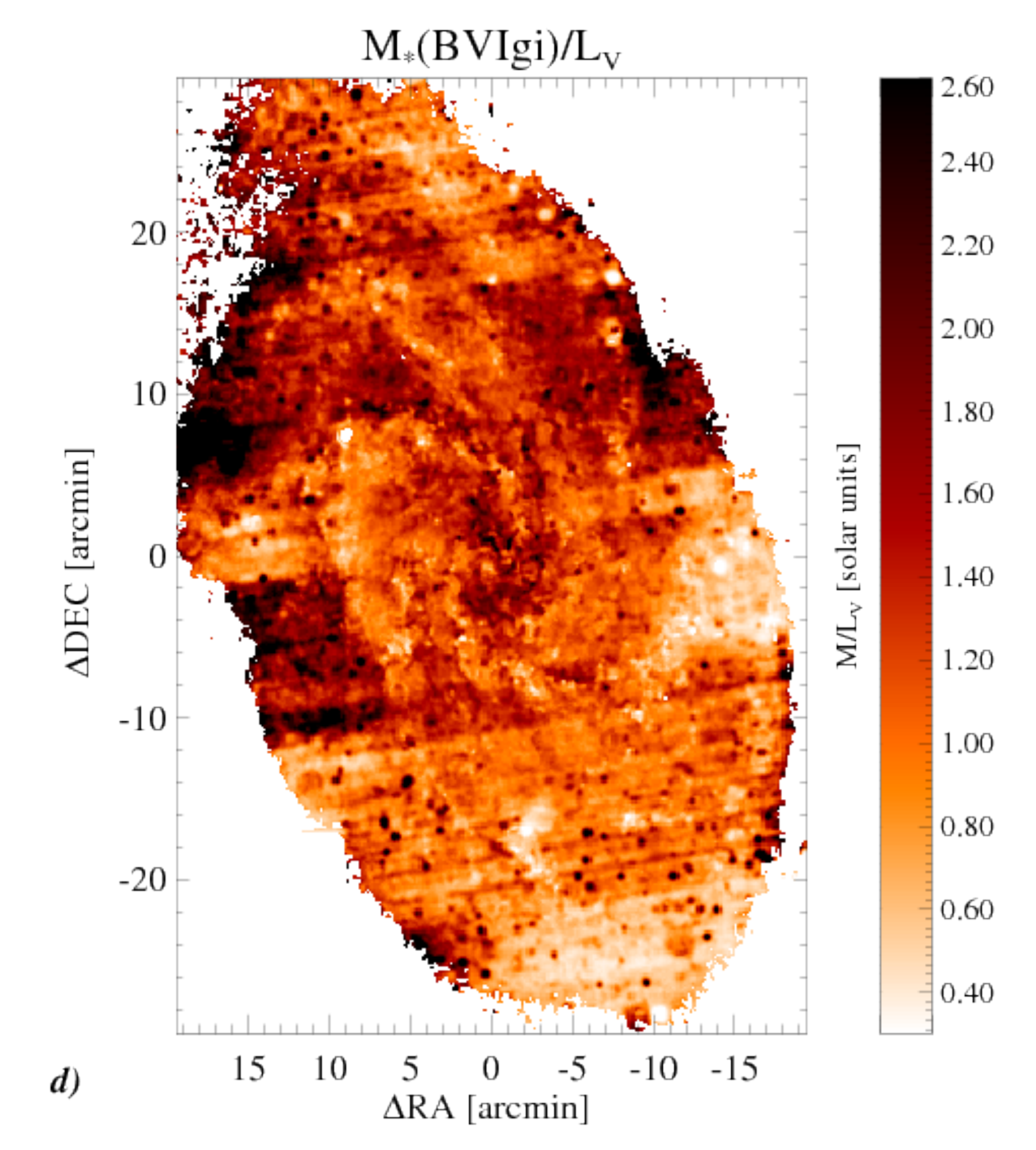}}
\caption{Panel \emph{a)}: The RGB color composite image of the B-V-I
  mosaics from the \citet{2006AJ....131.2478M} survey.  Panel
  \emph{b)}: Stellar mass map derived from the median likelihood estimator (described in \S\ref{mstar_method})
  based on LGS BVI bands. Panel \emph{c)}: Stellar mass map obtained from the bayesian median
  likelihood estimator, based on the full LGC and SDSS $BVIgi$  dataset. Panel \emph{d)}: The apparent stellar $M/L$ derived from
  the median likelihood stellar mass map based on LGS+SDSS $BVIgi$ bands
  and the observed LGS V-band surface brightness.}
\label{fig_optical} 
\end{figure*}

\subsection{The Local Group Survey}\label{mstar_LGS}
 
The Local Group Survey \citep{2006AJ....131.2478M} provides a set of
deep wide-field images of M33 in the $UBVRI$ passbands (only $B$, $V$ and $I$
will be used to estimate the stellar mass in the following), together
with a photometric catalog of 146,622 stars. Due to restrictions in
the field of view, the galaxy was covered by acquiring three partially
overlapping fields (C,N,S) in each band, a total of 15 images.  All
these images, together with those secured in narrow bands, are
publicly accessible
\footnote{ftp://ftp.lowell.edu/pub/massey/lgsurvey} but no photometric
calibration is provided due to the composite nature of the mosaicing
CCD detector.  In order to calibrate the images, we select  from the
photometric catalog a set of stars brighter than $V=16.00$: 195 stars
in the C field, 114 in the N, and 76 in the S. First we measure  the
size of the PSF in the images and then degraded all of them to match
the seeing of the worst
one (the N field in the $I$ band with a gaussian $\sigma=0.35$ arcsec).
Next, we perform multi-aperture photometry of the
selected stars and derive  the calibration by comparing fluxes with the
values reported in the catalog.  Note,  the multi-aperture photometry also
provides the background offsets between the 3 different fields in each
band.  After readjusting the background and the scale of the N and S
fields to C, the fields are combined to produce an image of the
entire galaxy which is later sky subtracted by using the most external
regions.  After registering the positions to the V-band reference, the
images are again calibrated with the same procedure outlined above
but including color terms. 
A full resolution RGB color composite image of the $B-V-I$ bands is shown in
Fig. \ref{fig_optical}, top left panel.

The $B$, $V$ and $I$ band calibrated maps are then edited, degraded in resolution
and resampled to match the SDSS images described in the previous
subsection. From the calibration procedure of these final maps,
again by comparison with the photometry of the stars with $V \leq 16.00$ in the
Local Group Survey catalog, we determine uncertainties
in the zero-point of 0.02. 0.10, and 0.18~mag in the $B$, $V$
and $I$ band, respectively.
These final maps are also used for determining the background
subtraction. 20 fields containing no stars, each 20x20 arcsec in extent,
were selected in the far NE and SW regions at the borders of the image;
these fields show minimal emission in all three bands.
In each bandpass, we assume the background to be the mean of the 20
median values, and the background uncertainty to be the $1\sigma$ of the sample
(not of the mean); this uncertainty amounts to 30\% of the surface
brightness at $\mu_B = 25.0, \mu_V=24.3, \mu_I=23.5$, respectively.

\subsection{Stellar mass map(s)}\label{mstar_maps}

The left bottom panel of Fig. \ref{fig_optical} shows the map of stellar
mass derived from the pixel-by-pixel analysis described in
\S\ref{mstar_method}, using both the $BVI$ maps from LGS and the $gi$ maps
from SDSS. The smoothness and azimuthal symmetry of the map is
remarkable, although, especially at low surface density values, one
can clearly see background residuals appearing as large scale ``spot''
fluctuations (coming from the LGS maps) and fluctuations along stripes
(coming from the SDSS scans). {\it Random} errors on this map (as derived from the inter-percentile 
range of the PDF described in Sec. \ref{mstar_method}) range from 0.06 to 0.1 dex (approximately 15 to 25\%), 
including the contribution of both photometric uncertainties and intrinsic model 
degeneracies between colors and M/L, but not systematics (see end of this section).

In the bottom right panel of Fig. \ref{fig_optical} we display the map
of the ratio between the estimated stellar mass and the V-band surface 
brightness (as measured) in solar units. This figure makes it
obvious that a simple rescaling of the surface brightness by a
constant mass-to-light ratio (as is usually done in the dynamical
analysis) is a very crude approximation to the real stellar mass
distribution. In particular, the M/L gradients apparent from this map
substantially affect the slope of the mass profile,  as discussed more
in detail in Section~6.

Figure \ref{fig_optical}, with its many apparent artefacts, also highlights the
limits of our photometry, mainly due to the difficulties of
obtaining an accurate background subtraction in such a large-scale
mosaic. On the other hand, the redundancy of our dataset in terms of
wavelength coverage reveals its importance to limit the impact of
systematic effects when we compare the mass surface density
obtained from  the full $BVIgi$ dataset
(bottom left panel of Fig. \ref{fig_optical}) to that obtained from the LGS
dataset alone (upper right panel). In particular, the comparison shows
that \emph{i)} the anomalously high surface density that appear in
the outskirts of the map based on LGS alone are corrected by including
also SDSS imaging, most likely thanks to a more accurate background
subtraction in the SDSS; \emph{ii)} the North-South asymmetry evident
in the LGS map is also corrected by a more stable and accurate
zero-point determination in the SDSS. Moreover, we note that the stripe structure
introduced by the background subtraction in the SDSS images is partly
removed by the inclusion of LGS images.

Despite some differences in the overall normalization, the mass surface density 
variations in the two  maps are consistent.   The stellar mass map derived from the full
dataset (LGS+SDSS) will be our  reference mass distribution but we shall carry out
the dynamical analysis using both maps, limiting the usage of the mass map from
the LGS dataset to R$<4$~kpc. Considering the uncertainties derived from our bayesian 
marginalization approach (thus combining model degeneracies and random photometric errors) and 
comparing the two mass maps obtained with the two photometric datasets, we estimate a 
typical stellar mass uncertainty of $\sim 30$\% (0.11 dex) out to $\mu_B=25$, which we apply to the dynamical analysis. 
We stress once again that the uncertainties derived from the interpercentile range of the PDF over a 
large and comprehensive library of model star-formation histories, metallicity and dust distributions, 
{\it de facto} realistically include the contribution of the systematic uncertainties related to the choice 
of models. This is one of the key advantage in using this approach with respect to other maximum-likelihood 
fitting algorithms, either parametric or non-parametric.

The only possible systematic uncertainties that are not included in the error budget are the ones related to the 
IMF, which is kept fixed in all models (to the Chabrier IMF), and those related to the basic ingredients of stellar 
population synthesis, i.e. the base SSPs. One can of course include models with different IMFs in the library, 
but broad band optical colors are essentially insensitive to it, hence they do not provide any constraint in 
that respect. Possible different IMFs should thus be treated as an extra freedom in the stellar mass normalization. 
Testing different SSPs is clearly out of the scope of this paper. However, it is worth mentioning that after many 
years of debate about the role of TP-AGB stars in stellar populations, the community is reaching a broad consensus 
on their limited role \citep[e.g.][]{kriek+10,conroy_gunn,zibetti+13}, thus leaving little freedom on the 
intrinsic M/L of simple stellar populations at fixed SED. Therefore this contribution to the mass error budget 
appears negligible in comparison with the variations induced by different star formation histories, metallicity 
and dust distributions, already accounted for by our method of error estimates. 

\subsection{The outermost stellar disk}\label{mstar_extern}

As pointed out by several surveys \citep[e.g.][]{2010ApJ...723.1038M}, the stellar disk of M33 extends out
to the edge of the HI disk following a similar warped structure. To account for this fainter but extended 
disk we first extrapolate the stellar mass surface density of the inner stellar map outward to 10~kpc, using 
the radial scalelength inferred by the 3.6~$\mu$m map \citep{2007A&A...476.1161V,2009A&A...493..453V}. This
scalelength, 1.8$\pm0.1$~kpc, is consistent with that inferred in our stellar surface density maps beyond R$\sim$ 2~kpc.
Stellar counts from deep observations of several fields further out, 
in the outer disk of M33, have been carried out by \citet{2011A&A...533A..91G} using the Subaru telescope.  
The typical face-on radial scalelength of the stellar mass density in the outer disk inferred from these observations is
25~arcmin ($\sim$6~kpc), much larger than that of the brighter inner disk and very similar to the HI scalelength in the
same region. We shall use this radial scaling to extrapolate  exponentially the
stellar surface density further out and truncate the  profile  with a sharp
fall off beyond 22~kpc.

\section{The rotation curve}

In this Section we outline the derivation of the rotation curve of M33 starting from an analysis of the
spatial orientation of the disk.  

\subsection{The warped disk}

To make a dynamical mass model of a disk galaxy it is necessary to  reconstruct the three-dimensional velocity field 
from the velocities observed along the line of sight. If velocities are circular and confined to a disk one  
needs to establish the disk orientation i.e. the position angle of the major 
axis (PA), and the inclination ($i$) of the disk with respect to the line of sight. If the disk exhibits a warp 
these parameters vary with galactocentric radius. This is often the case for gaseous disks 
which extend outside the optical radius and which  show different orientations than the inner ones. 
Tilted ring model are considered to infer the rotation curve of warped galactic disks.
\citet{1997ApJ...479..244C} have fitted a tilted-ring model to the 21-cm line data over the full extent of 
the M33-HI disk. However the disk was not fully sampled by the data
since the observations (carried out with the 
Arecibo flat feed, FWHM=3.9~arcmin) only sampled the disk over an hexagonal grid with 4.5~arcmin spacing.  
To determine the disk orientation using the new 21-cm all-disk survey
described in this paper, we follow a  tilted ring
model procedure 
which departs from the usual schemes and which has been used by \citet{1997ApJ...479..244C}
and  later  implemented by \citet{2007A&A...468..731J} (TiRiFiC, a Tilted Ring Fitting Code, available
to the public). A set of free rings is considered, each  ring being characterized by its radius $R$ and by 7 
additional parameters: the HI surface density $S_{HI}$, the circular velocity $V_c$, the inclination 
$i$ and the position angle PA, the systemic velocity $V_{sys}$ and 
the position and velocity shifts of the orbital centers with respect to the galaxy center
($\Delta x_c, \Delta y_c$). 
Importantly, rather than fitting  the moments of the flux distribution, we infer the best fitting parameters by comparing the 
synthetic spectra of the tilted ring model to the full spectral database, as explained in details in Appendix~\ref{appa}. In 
Appendix~\ref{appa} we described also the two minimization methods used,  the 'shape' and 
'v-mean' methods, which give consistent results for the M33 ring inclinations and position angles.

\begin{figure} 
\includegraphics[width=\columnwidth]{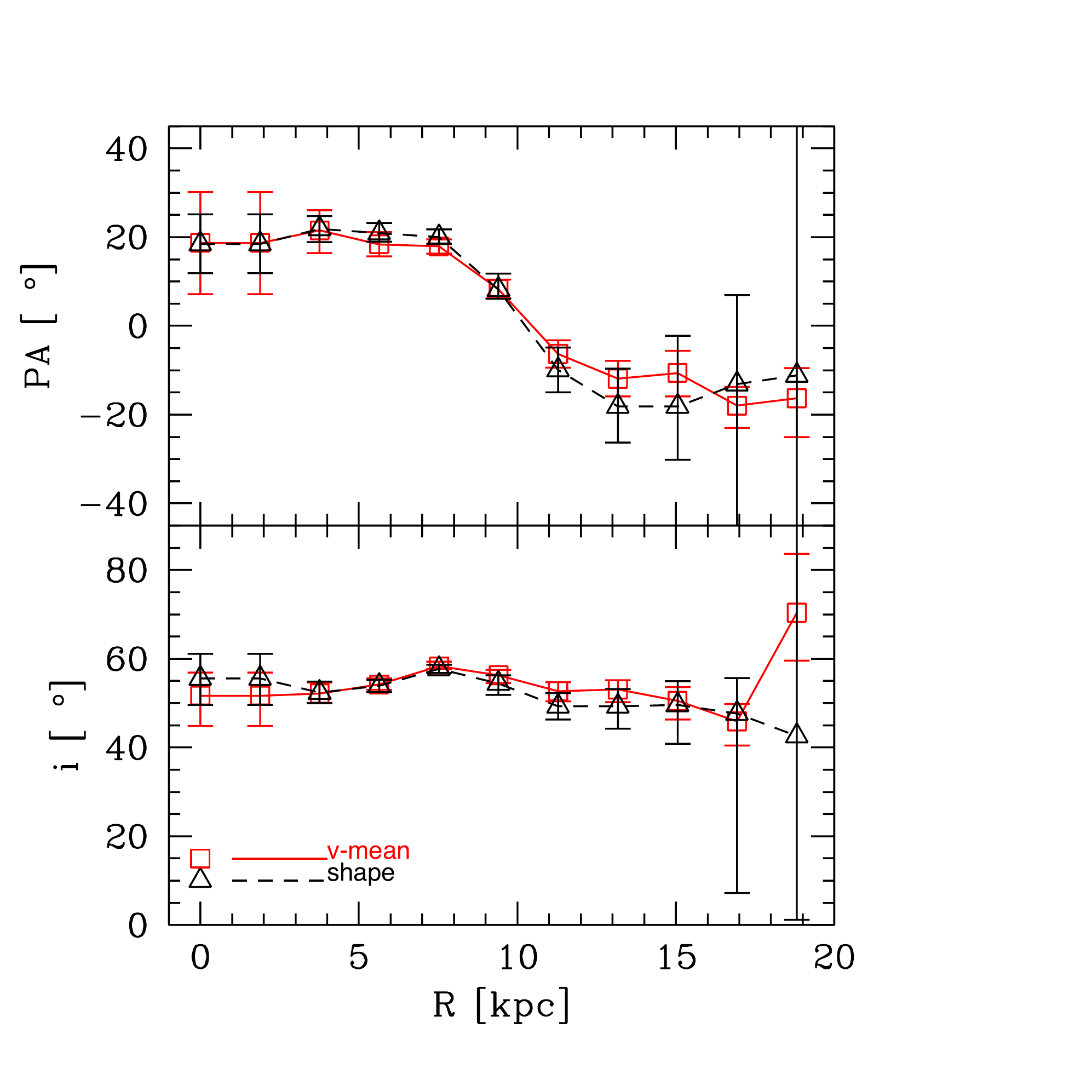} 
\caption{The inclination and position angle of the 11 free tilted rings used for deconvolving the 21-cm data.
The open triangles (in black in the on-line version), connected by a dashed line, shows the value using the shape minimization 
method, the open squares (in red in the on-line version),
connected by a continuous line, are relative to the v-mean minimization method (see Appendix~\ref{appa} for details). 
The uncertainties are computed by varying one parameter of one free ring  at a time  around the minimal solution 
until the $\chi^2$ of the tilted ring model increases by 2$\%$ .}
\label{pa-i} 
\end{figure}

We display the best fitting values of $i$ and PA in Figure~\ref{pa-i} with their relative uncertainties. 
As we can see from Figure~\ref{pa-i}  the orientation of the outermost rings has high uncertainties. 
This is especially remarkable when using the "shape" method and hence  we fix the outermost 
ring parameters to be equal to those of second-last ring for deconvolving the data. The intermediate regions solutions  
appear to be more robust. 
In the next subsection the best fitting values of systemic velocity shifts and ring center displacements 
(shown in Appendix~\ref{appa}) will be used together with
the rings orientation angles, $i$ and PA, to derive  rotational velocities V$_r$ and the face-on surface brightness 
$\Sigma_{HI}$ from the data at high and low resolution. Notice that the values of V$_r$ and 
$\Sigma_{HI}$ we infer are consistent with the value of the free  parameters 
$V_c$ and $S_{HI}$ of the adopted ring model but will be sampled at a higher resolution along the radial direction.

\subsection{The 21-cm velocity indicators}

To derive the rotation curve we use the 21-cm datacube at a spatial resolution of 20 and 130~arcsec.
Emission in the high spatial resolution dataset is visible only out to R=10~kpc while in the lower resolution 
dataset  the HI is clearly visible over a much extended area.
To trace the disk rotation we consider both the peak and the flux weighted mean velocities (moment-1) along the line of sight 
of the 21-cm line emission at the original spectral resolution.
The velocity at the peak  of the line is in general a better indicator of the disk rotation than the mean velocity when the 
signal-to-noise is high and the rotation curve is rising  \citep[e.g.][]{2008AJ....136.2648D}  and we will use
this velocity indicator inside the optical disk. 
In the outer disk, where the signal-to-noise is lower and the rotation curve is flatter we  use instead the 
flux weighted mean velocities  (as well shall see, the curves traced by the two velocity indicators in this regions are
quite consistent).  

The rotation curve of M33 is extracted from the adopted line-of-sight velocity indicator using as a
deconvolution model the best-fit disk geometrical parameters (derived via the shape and v-mean methods, described in 
Appendix~\ref{appa}, which we shall call model-shape  and  model-mean). The model-shape and model-mean parameters are  
shown in Figure~\ref{ringmodels} and in Figure~\ref{pa-i}. If  21-cm emission is present in some area located at larger 
radii than the outermost free ring  
we deconvolve the observed velocities assuming a galaxy disk orientation as the outermost free ring. 
  
When using the mean instead of the peak  velocity, the same deconvolution
model results in a rotation curve with lower rotational velocities.
In the optical disk the difference between the rotation curve traced by the peak or the mean velocities can be as 
high as 10~km~s$^{-1}$ but the scatter between the results relative to the two deconvolution models is small 
(it is always less than $2$~km~s$^{-1}$ and for most bins less than $1$~km~s$^{-1}$). Further out there are no systematic 
differences between the curve traced by the peak and the mean velocity but the outermost part is very sensitive to the
orientation parameters adopted. These parameters  are not tightly constrained in the outermost regions
due to the partial coverage of the modeled rings with detectable 21-cm emission and to some multiple component spectra.
Here it will be important to consider the deconvolution model uncertainties in the rotation curve. 
 
\begin{figure} 
\includegraphics[width=\columnwidth]{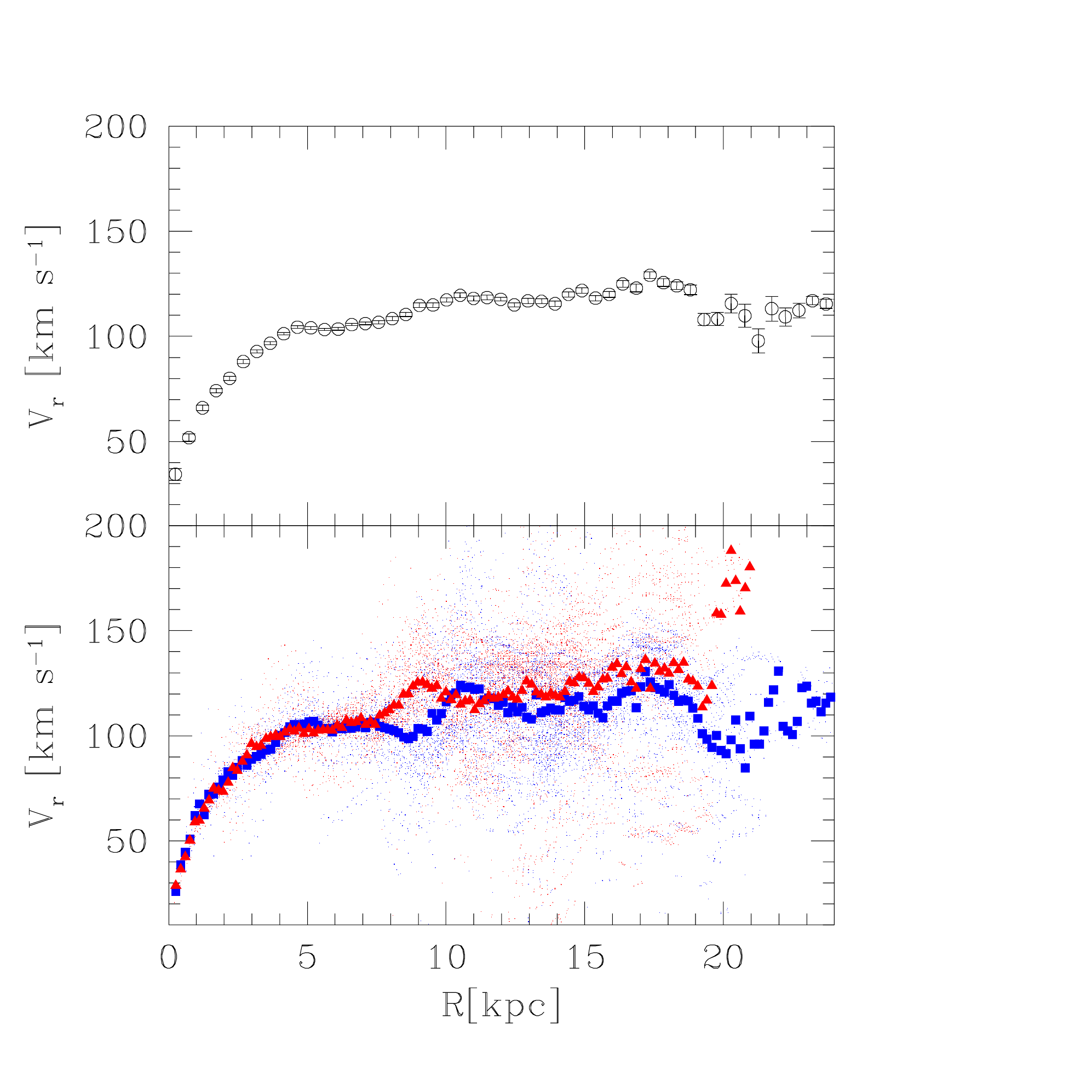} 
\caption{The bottom panel shows with small dots the rotational velocities for each pixel in the peak map 
 (red dots for the southern half and blue dots for the northern half in the on-line version) 
and with large symbols the average values in each radial bin. The model-shape has been used to deconvolve the data.
Triangles (red in the on-line version) are for the southern half while squares
(blue in the on-line version) are for the northern half. The top panel shows the average rotation curve for  
peak velocities.  Bins  are 2~arcmin wide in the radial direction.}
\label{peak-shape} 
\end{figure} 

\begin{figure} 
\includegraphics[width=\columnwidth]{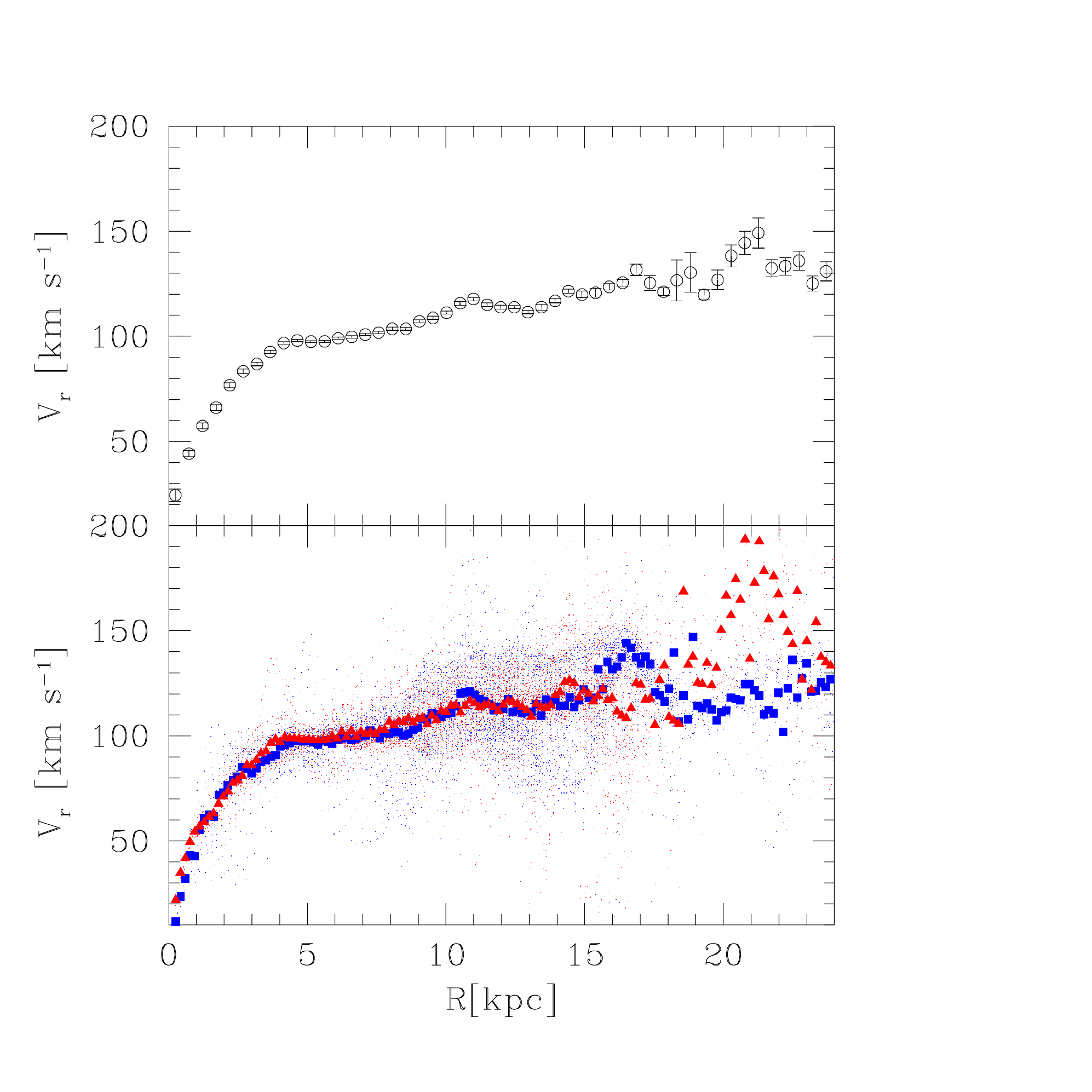} 
\caption{The bottom panel shows with small dots the rotational velocities for each pixel in the first moment-1 map 
and with large symbols the average values in each rings   (Red symbols are for the southern half and blue symbols 
for the northern half in the on-line version). The model-mean has been used to deconvolve the data.
Triangles are for the southern half while squares are for the northern half. The top panel shows the average rotation curve for  
moment-1 (mean) velocities.  Bins  are 2~arcmin wide in the radial direction. }
\label{mom1-mean} 
\end{figure} 

\begin{figure} 
\includegraphics[width=\columnwidth]{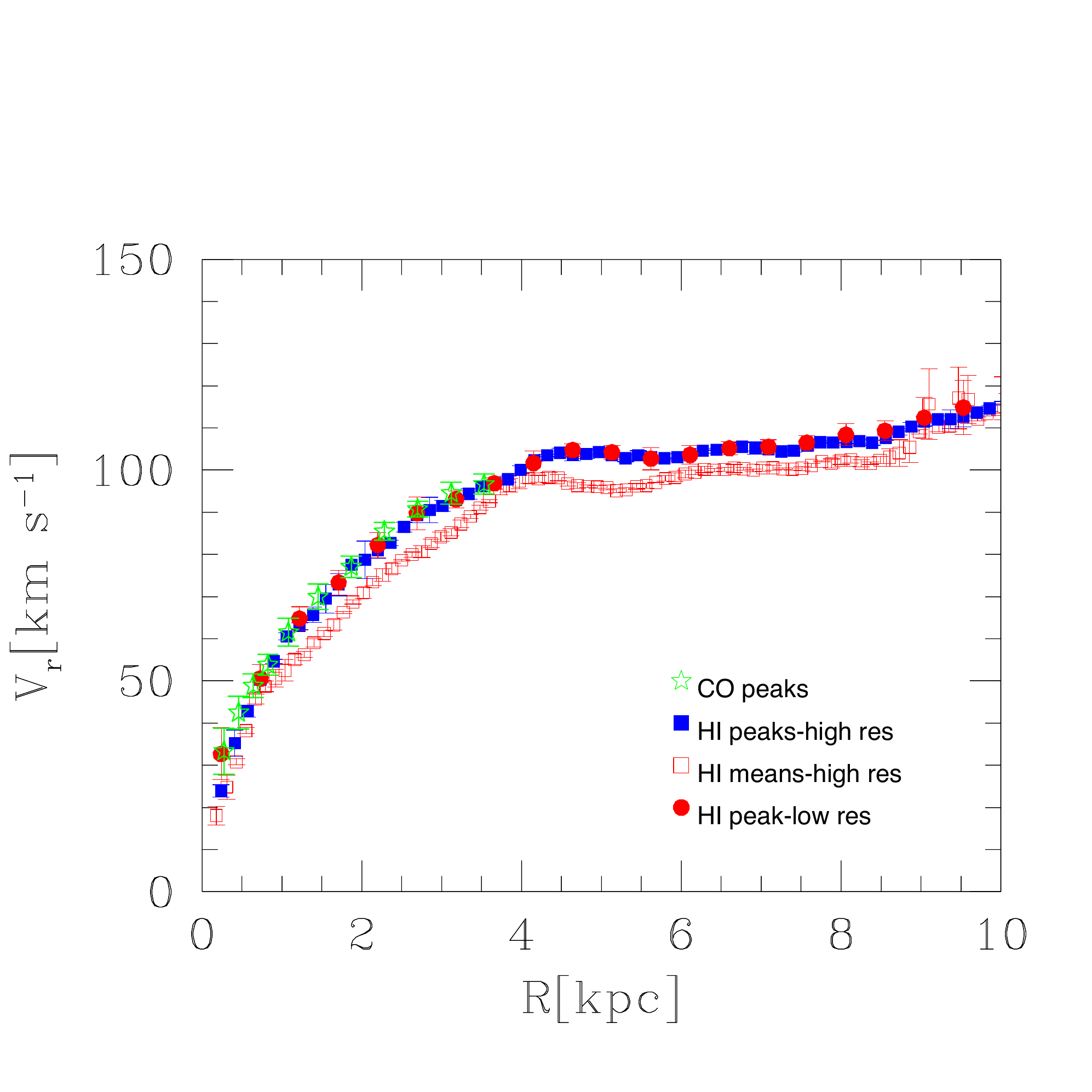} 
\caption{The inner rotation curve as traced by the peak velocities at high resolution (filled squares  in blue in the
on-line version), 
and at low resolution (filled dots  in red in the on-line version). The CO data are shown with star symbols (in green 
in the on-line version). The open square  symbols (in red in the on-line version) are for
rotational velocities as traced by the moment-1 map.  For each curve the weighted mean velocities between that 
relative to deconvolution model-shape and model-mean are used. }
\label{mom1-peak} 
\end{figure}

In Figure~\ref{peak-shape} and in Figure~\ref{mom1-mean} we show the rotation curve from the 
peak and the mean velocities, the first one obtained from the
tilted-ring model-shape and the second one from model-mean.
For the outer disk  we use the low resolution dataset with 2~arcmin radial bins (0.5~kpc wide).
For the inner disk the high resolution data are binned  in 40~arcsec radial bins (163~pc wide). 
In Figure~\ref{mom1-peak} we show the curve traced by the peak velocities after averaging the rotational velocities
relative to the deconvolution model-shape  and deconvolution model-mean.  In the same Figure the filled dots
show the curve from the low resolution dataset which
matches perfectly the high resolution curve. To not overweight the inner rotation curve in the dynamical analysis we will use
the low resolution 21-cm dataset in addition to the velocities of the CO-J=1-0 line peaks (black stars in Figure~\ref{mom1-peak}).
The  average values of PA and $i$  between deconvolution model-shape and model-mean are used to trace the rotation curve with 
CO J=1-0 line. These deconvolution parameters have been used also to trace  the inner curve with the moment-1 
velocities, which is shown for comparison in  Figure~\ref{mom1-peak}  (open square symbols).  There is
not much difference between the rotational velocities retrieved from model-shape or model-mean in the inner region but the
moment-1 velocities give clearly lower rotation curve than the peak velocities.

\subsection{Rotation curve: a comparative approach}

\begin{figure} 
\includegraphics[width=\columnwidth]{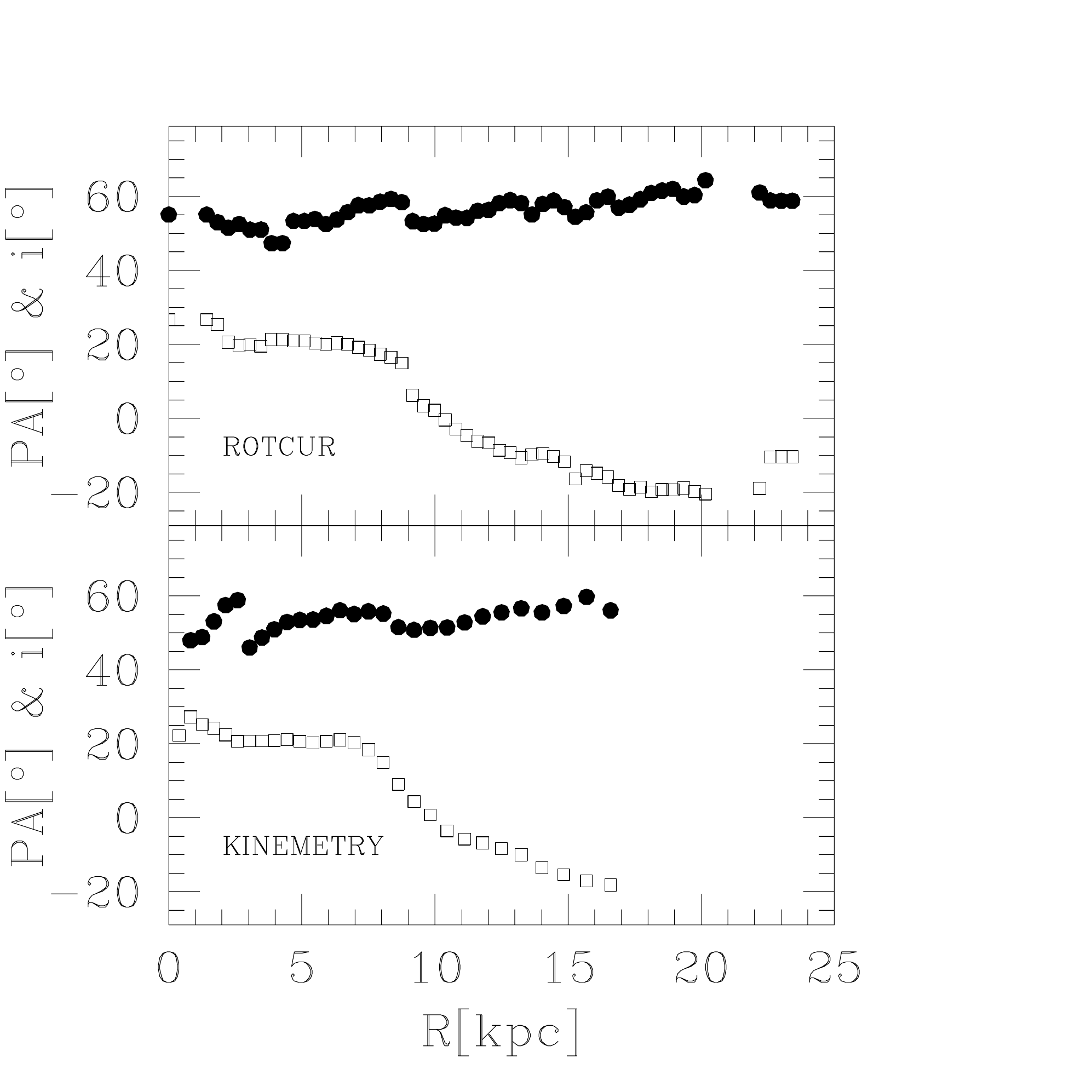} 
\caption{The tilted ring position angles and inclinations are shown with open square symbols and filled dot 
symbols respectively for the ROTCUR and KINEMETRY fitting routines.
 }
\label{nemokine} 
\end{figure}

\begin{figure} 
\includegraphics[width=\columnwidth]{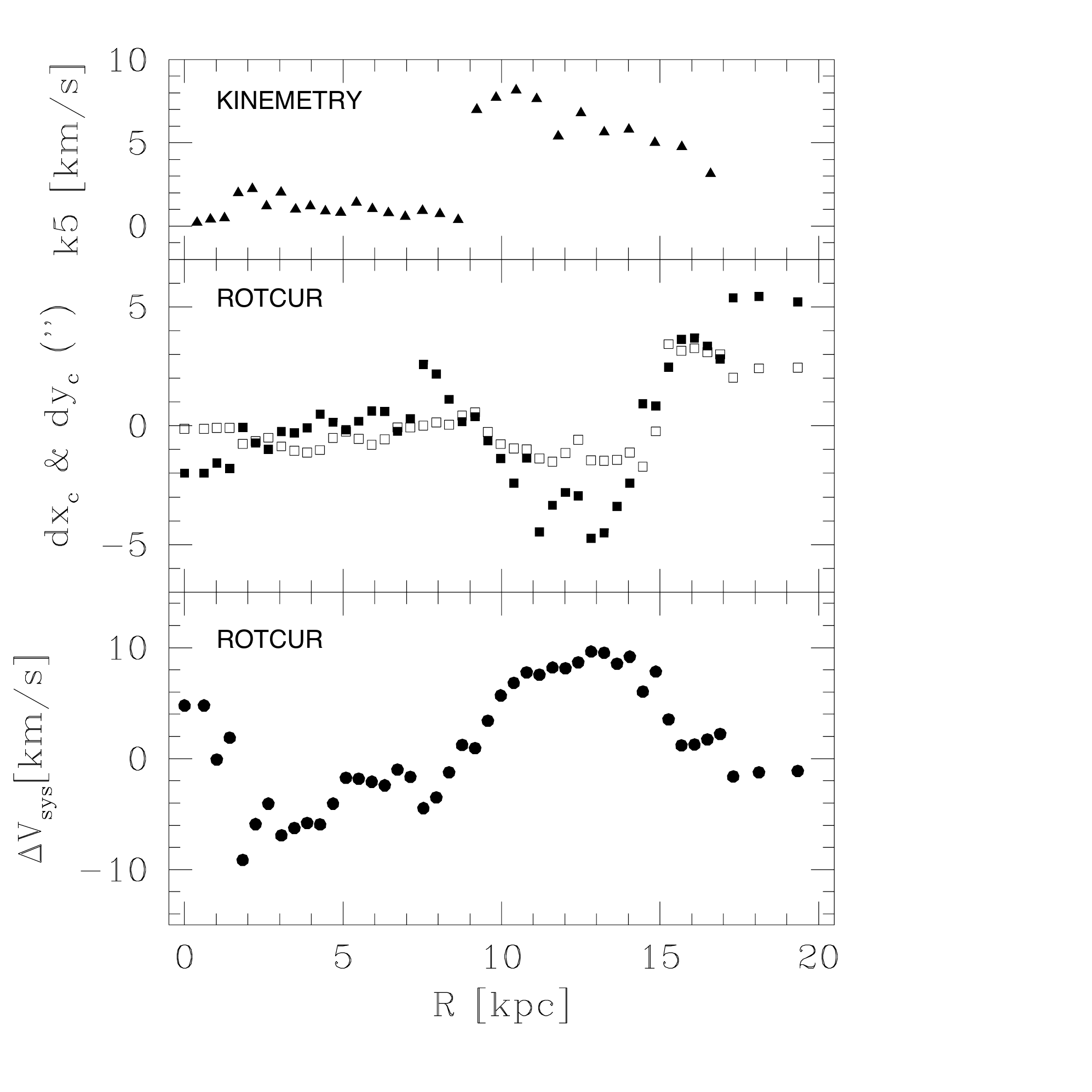} 
\caption{Some of the tilted ring parameters fitted by the routines ROTCUR and KINEMETRY. The top panel
shows the term k5, which is an higher order term  in the  KINEMETRY outputs. In the bottom panel we show the shifts
of the systemic velocity and in the middle panel the orbital center shifts along the y-axis in the South-North direction 
(filled square symbols) and  along the x-axis oriented from East-to-West (open square symbols).}
\label{shift} 
\end{figure}

In this subsection we compare the rotation curve derived through our technique of fitting the 21-cm line emission    
(the datacube) with a tilted ring model, to those resulting from other type of approaches and methods.
These complementary approaches are useful to strengthen the robustness of the rotation curve and  to better define  
the uncertainties. We consider two additional methods devised for deriving galaxy rotation curves from  two-dimensional 
moment-maps.
The first one is the standard least-square fitting technique developed by \citet{1987PhDT.......199B}  as implemented in the 
ROTCUR task within 
the NEMO software package  of analysis \citep{1995ASPC...77..398T}. The other method is based on the harmonic decomposition 
of the velocity
field along ellipses and we  use for this purpose the software KINEMETRY developed and provided by \citet{2006MNRAS.366..787K}.
Both methods work on 2D momentum maps i.e. on one velocity per pixel, and  minimize the free parameters of one ring at a time,
without accounting for possible overlaps of ring pieces in the beam. We run ROTCUR in two steps: at first we let the ring 
centers and V$_{sys}$ vary.  The orbital center shifts are larger in the 
outermost regions and roughly in the direction of M31, as found by \citet{1997ApJ...479..244C}, and are shown in 
Figure~\ref{shift}. In the second iteration we fix the ring centers and V$_{sys}$ to the average values found in the first 
iteration, x$_c$=-33~arcsec  y$_c$=105~arcsec  V$_{sys}$=-178~km~s$^{-1}$. We run a second iteration because by 
fixing the orbital centers the routines converge  
further out, at  radii as large as R=23.5~kpc. In this second attempt we run
ROTCUR with 51 free rings, uniform weight, excluding data within a 20$^\circ$ around the minor axis. The resulting PA and $i$
are shown in the top panel of Figure~\ref{nemokine}.
With the KINEMETRY routines we fit with a smaller number of free rings, 10, and convergence is achieved over a smaller radial range,
out to R=17.5~kpc. The higher order terms of the  harmonic decomposition are useful tools if one is looking for non-circular 
motion in  the disk. In the bottom panel of Figure~\ref{nemokine} we show the usual PA and $i$  but in addition we plot the
coefficient $k_5$ in Figure~\ref{shift}. This is an higher order term  which represents deviations
from simple rotation due to a separate kinematic component like radial infall or non-circular motion. 
Its amplitude is negligible in the optical disk but it is  as high as 7~km~s$^{-1}$ in the outer disk. If the anomalous velocities
are in the radial direction this gas can be fueling star formation in the inner disk.

\begin{figure} 
\includegraphics[width=\columnwidth]{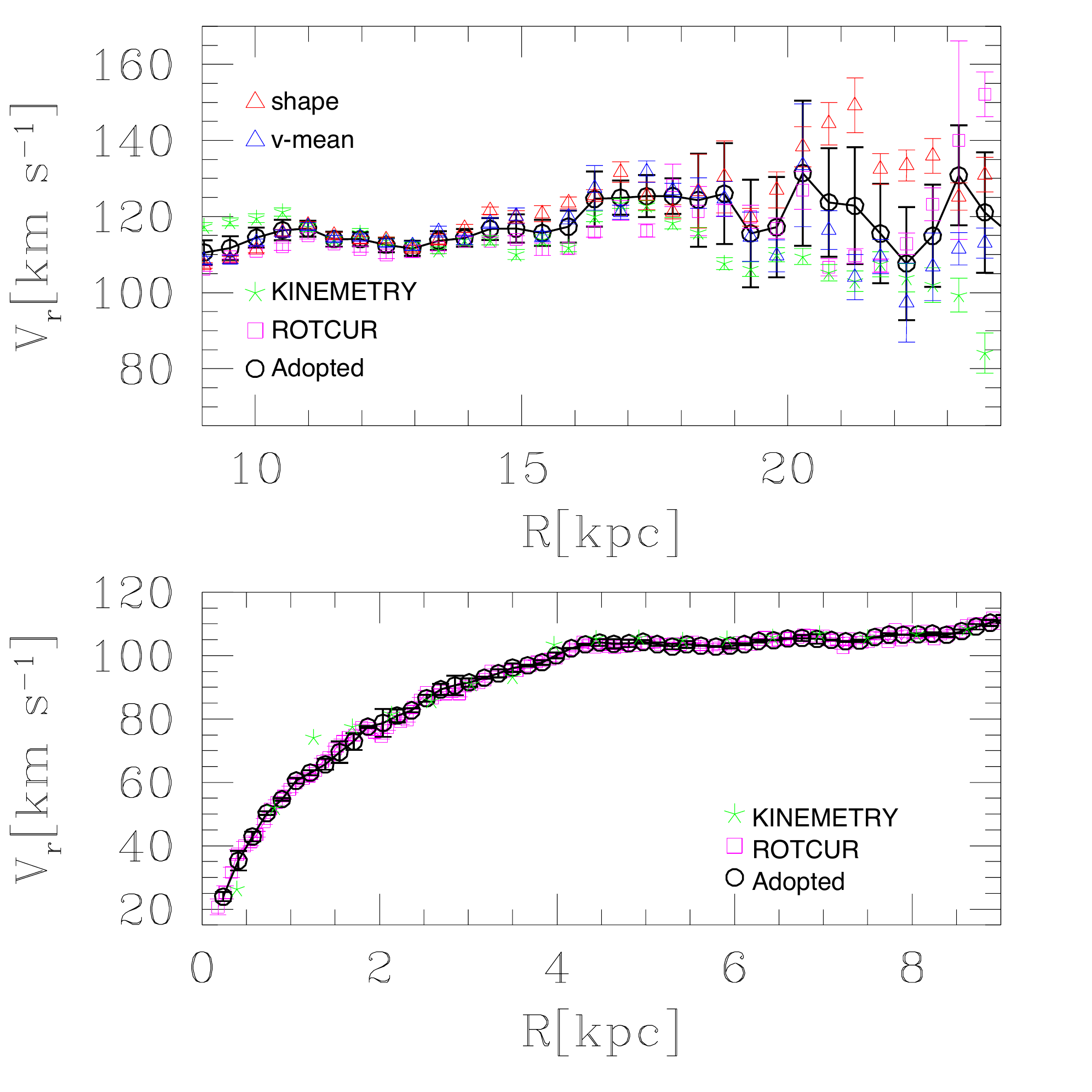} 
\caption{The open black circles show the adopted rotation curve for the inner regions (bottom panel) using the peak velocities
and for the outer regions (top panel) using the first moments. The dispersions take into account variation of V$_r$ 
between different deconvolution models,
the 2$\%$ variations around the minimum $\chi^2$ and the standard deviation of the mean velocity in each bin.
In both panels the open square and asterix symbols  (in magenta and green  in the on-line version) 
show  V$_r$ according to the ROTCUR and KINEMETRY
deconvolution model respectively. In the top panel the open triangles are for the first moment velocities
deconvolved according to  model-shape and model-mean  
(in red and blue colors respectively in the on-line version).}
\label{compa} 
\end{figure}

The two rotation curves obtained with the NEMO and KINEMETRY routines are very similar to the one derived from 
the deconvolution model-shape and -mean inside the optical disk, but there are some differences in the resulting warp orientation 
and hence in the outer rotation curve. This can be seen  in Figure~\ref{compa}. There are very marginal differences between
the rotation curves of the inner region relative to  different deconvolution models. Beyond 15~kpc instead there are
significant differences which we take into account.

\subsection{The adopted rotation curve and its uncertanties}

For the final rotation curve of M33 we use the following recipe. 
Inside the optical disk we use representative  
21-cm peak circular velocities computed as the weighted mean of velocities
originating from our two deconvolution choices
(model-shape and model-mean). To these we append the CO dataset. Beyond 9~kpc, in each radial bin
we use the weighted mean circular velocity computed by averaging the moment-1 velocities resulting from: (1) the deconvolution
model-shape, (2) the deconvolution model-mean, (3) the package KINEMETRY
and (4) the task ROTCUR in NEMO package.

The rotation curve uncertainties take into account not only the standard deviations around the mean  of the
2 or 4 velocities  averaged, but also the data dispersion in each radial bin (see data in bottom panels of Fig~4 and 5)
and the uncertainties relative to deconvolution models ($\Delta V_r$ corresponding to 2$\%$ $\chi^2$ variations of 
the tilted ring model-shape and -mean  through PA and $i$ displacements).

To the final rotation curve we  apply the small finite disk thickness corrections, described in Appendix~\ref{appb}.

\section{The surface density of the baryons}

\begin{figure} 
\includegraphics[width=\columnwidth]{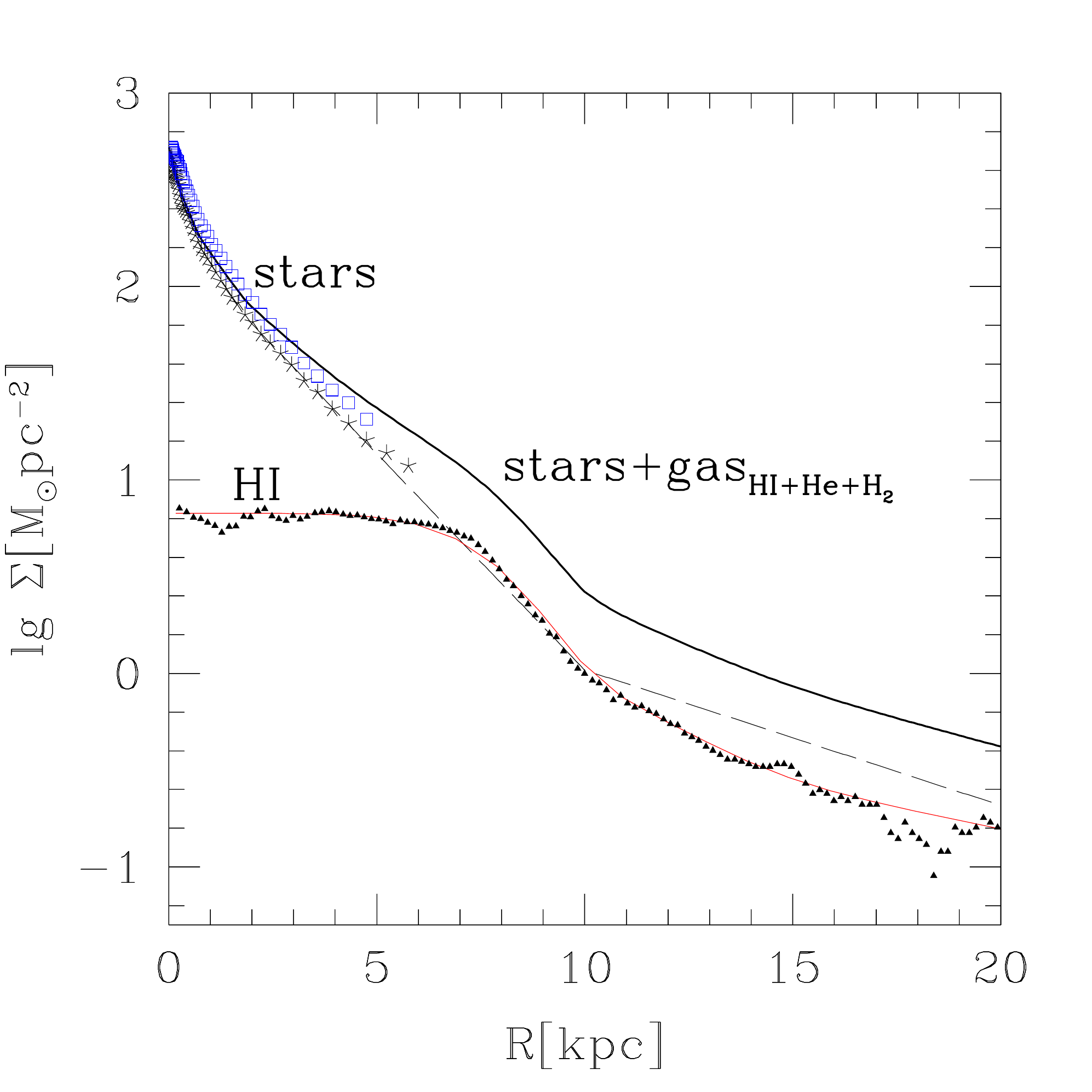} 
\caption{The HI surface density perpendicular to the galactic plane of M33 (small filled triangles) 
and the function which fits the data (continuous line, {\bf red in the on-line version}) after the 21-cm line 
intensity has been deconvolved 
according to tilted ring model-shape. Asterix symbols indicate the stellar mass surface density using the $BVIgi$
stellar surface density map. The dashed line is the fit to the stellar surface density and the extrapolation to
larger radii. Open squares {\bf (in blue in the on-line version)} show for comparison the surface density using the 
BVI mass map. The heavy weighted
line is the total baryonic surface density, the sum of atomic and molecular hydrogen, helium and stellar mass
surface density.}
\label{surface} 
\end{figure}

In this  Section we use the radial profile of the stellar mass surface density  
from the $BVIgi$ maps, together with the gaseous surface density, to compute the total baryonic 
face-on  surface mass density of M33 and the stellar mass-to-light ratio.
The surface density of stars perpendicular to the disk is shown in Figure~\ref{surface}:
it drops by more than 3 orders of magnitudes from the center to the  
outskirts of the M33 disk. The total stellar mass according to the $BVIgi$-mass map extrapolated
to the outer disk is 4.9
10$^9$~M$_\odot$ (5.5~10$^9$~M$_\odot$ for the BVI map) of which about 12~$\%$ resides in the outer disk.
At the edge of the optical disk \citet{2011MNRAS.410..504B} find a surface density of stars $\sim$1.7~M$_\odot$/pc$^2$
for a Chabrier IMF down to 0.1 solar masses assuming the inclination of our tilted ring model. 
This is consistent to what our modeled radial stellar distribution predicts at 9~kpc: 1.7$\pm 0.5$~M$_\odot$/pc$^2$.
The outermost field (S2) of \citet{2011MNRAS.410..504B}  
has been placed outside the warped outer disk and hence traces only a possible stellar halo. 

Using the best fitting tilted ring model we derive the radial distribution of neutral atomic gas, 
perpendicular to the galactic plane, in the optically thin approximation. This is shown in Figure~\ref{surface}. 
The total HI mass computed by integrating the
surface density distribution in Figure~\ref{surface} out to 23~kpc is about 20$\%$ higher than the true
HI mass of the galaxy, which is 1.53$\times 10^9$~M$_\odot$. This is because we average the flux of all pixels 
with non-zero flux in each ring and the HI emission in the outermost rings 
does not cover the whole ring surface. We do this because it is likely that  undetected pixels at 21-cm   
are not empty areas  but host ionized gas. In fact, a sharp HI edge has been detected in this galaxy 
as the gas column density approaches 2$\times$ 10$^{19}$~cm$^{-2}$ and interpreted
as an HI-->HII transition  \citep{1993ApJ...419..104C}. 
Considering the irregular outer HI contours of M33 as being due to ionization effects,
the outermost part of the atomic radial profile is in reality an HI+HII profile since there is HII where HI lacks 
at a similar column density. This assumption has a negligible effect on the
dynamical analysis of the rotation curve.  
The continuous line in Figure~\ref{surface} is the log of the function used to compute the dynamical
contribution of the hydrogen mass to the rotation curve. The total baryonic surface density is
 computed adding to the hydrogen gas the stellar mass surface density, the molecular and
the helium mass surface density (as given in Section 2.2).  Stars dominate the potential in the
star-forming disk ($R<7$~kpc), beyond this radius stars and gas give a similar contribution to the baryonic
mass surface density and decline radially in the outer disk with a similar scalelength.  
 
\begin{figure} 
\includegraphics[width=\columnwidth]{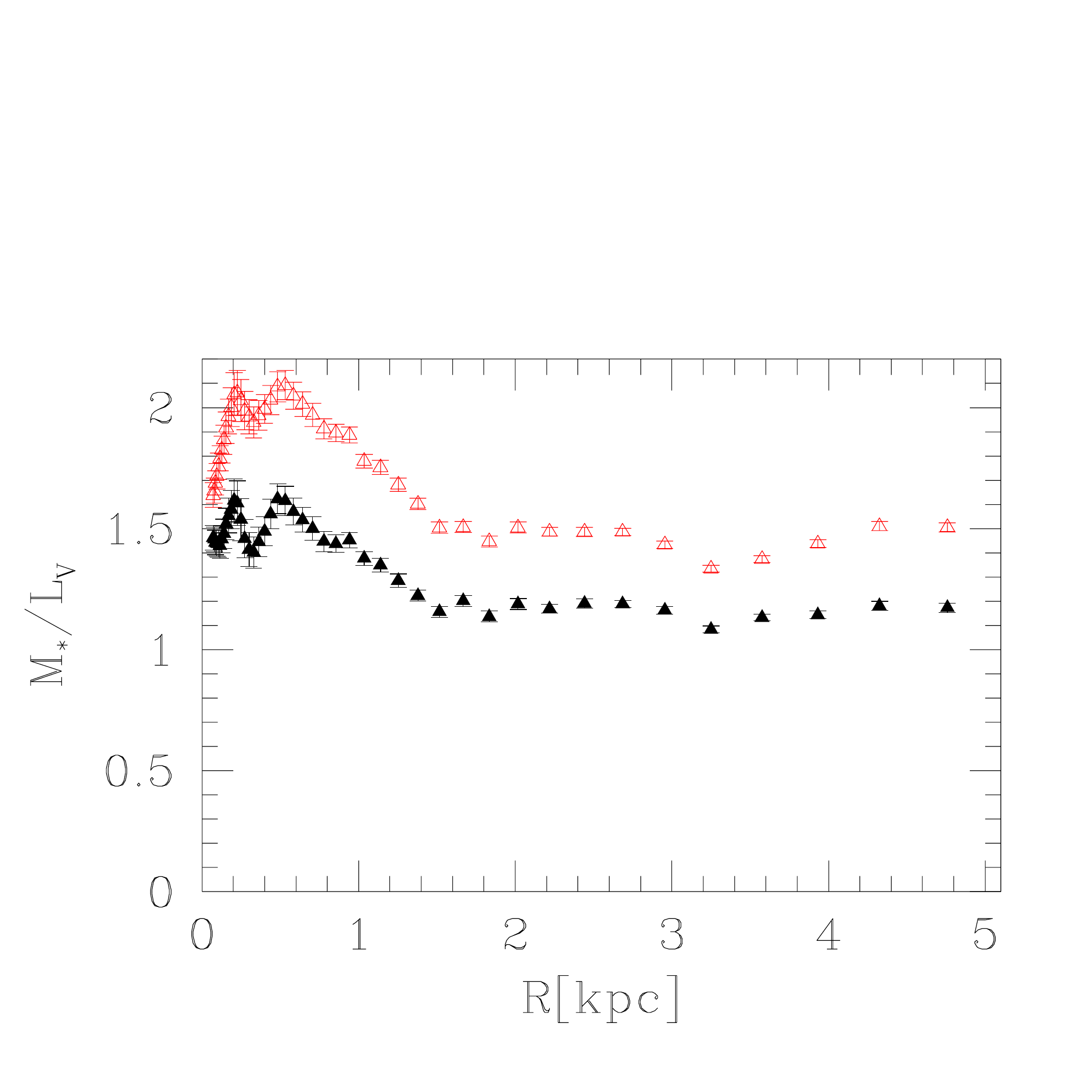} 
\caption{The mass-to-light ratio as a function of galactocentric radius according to the $BVIgi$ stellar surface 
density map (filled triangles, in black in the on-line version). This is computed by averaging the stellar
mass map and the V-band surface brightness along ellipses corresponding to  
the tilted ring model. We then compute  M$_*$/L$_V$ as the ratio  between these two quantities in units of 
M$_\odot$/L$_\odot$. We show for comparison the same ratio relative to the BVI stellar surface density map
(open triangles, in red in the on-line version). The error bars are  the standard deviations relative to
azimuthal averages and do not take into account uncertainties in the mass determination.}
\label{ml} 
\end{figure} 

The mass-to-light ratio as a function of galactocentric radius  is computed by averaging the $BVIgi$ stellar
surface density map and the V-band surface brightness along ellipses corresponding to  
the tilted ring model. The filled triangles in Figure~\ref{ml} show the ratio between these quantities, M$_*$/L$_V$,  
in units of M$_\odot$/L$_\odot$. We show for comparison the same ratio relative to the BVI stellar surface density map
(open triangles). Despite some differences in the two maps, the shape of the azimuthally averaged 
radial profiles of the mass surface density are consistent. Only in the innermost 1~kpc there is some difference
in the radial decline. The error bars are  the standard deviations relative to
azimuthal averages and do not take into account uncertainties in the mass determination. We clearly see radial
variations of the mass-to light ratio, especially in the innermost 1.5~kpc relative to areas at larger galactocentric radii.
This is an effect of the inside-out disk formation and evolution which adds to more localized variations
present in the map such as in arm-interarm contrast.

Average values or radial dependencies of the mass-to-light ratios, shown in Figure~\ref{ml}, can be compared with those 
derived previously using different methods.  The dynamical analysis of planetary nebulae in M33 \citep{2004ApJ...614..167C} gives a 
mass-to-light ratios which increase radially, being 0.2~M$_\odot/L_V\odot$ at R=2~kpc and 0.8~M$_\odot/L_V\odot$ at 5~kpc.
However, this result is based on the assumption of radially constant vertical scaleheight and of a much longer  
radial scalelength of the stellar surface density than suggested by near-infrared photometry. These assumptions imply also
a strong radial decrease of the velocity dispersion if the disk is close to marginal stability, in disagreement with
what is observed for the atomic gas. A more recent revised model for the marginally stable  disk of M33 \citep{2012AstL...38..139S}implies
a flatter mass-to-light ratio, of order 2~M$_\odot/L_V\odot$ in closer agreement with what we find. An increase of the mass-to-light
ratio in the central regions of M33 and a stellar mass surface density with  very similar values to what we show in 
Figure~\ref{surface} has been found by \citet{2009ApJ...695L..15W} using resolved stellar photometry and modeling of
the color-magnitude diagrams.

The former M33 rotation curve \citep{2003MNRAS.342..199C} was best fitted using CDM dark halo models  
by assuming an average stellar mass-to-light ratio in the range 0.5-0.9~M$_\odot/L_V\odot$, a somewhat 
smaller value that derived here. However \citet{2003MNRAS.342..199C} considered an additional bulge component 
whose presence has not been supported by subsequent analysis \citet{2007ApJ...669..315C}. From the stellar mass model
presented here it seems more likely that the M33 disk  has larger mass-to-light ratio in the innermost 1.5~kpc
rather than a genuine bulge. As it will be shown in the next Section, a radially varying mass-to-light ratio without a bulge 
component gives a similar CDM halo model for M33.  On the other hand the
constraints on the stellar mass-to-light ratio given by our mass map will be hard to reconcile with some  dark halo model.

\section{Tracing dark matter via dynamical analysis of the rotation curve}

The dynamical analysis of a high resolution rotation curve, such as that presented in this paper for M33, together with
detailed maps of the baryonic mass components, provides a unique test for the dark matter halo density
and theoretical models of structure formation and evolution. In this Section we 
shall consider a spherical halo with a dark matter density profile as 
originally derived by \citet{1996ApJ...462..563N,1997ApJ...490..493N} 
(hereafter NFW) for galaxies forming in a Cold Dark Matter scenario. We 
consider also the Burkert dark matter density profile (or core model) 
\citep{1995ApJ...447L..25B} since this successfully fitted the rotation  
curve of dark matter dominated dwarf galaxies \citep[e.g.][ and references 
therein]{2007MNRAS.375..199G}. 
Both models  describe the dark matter halo density profile using two  
parameters which we   determine through the best fit to the rotation curve.
Our last attempt will be to fit the rotation curve using MOdified
Newtonian Dynamics (MOND).
 
We perform a dynamical analysis of the rotation curve in the radial range: $0.4\le R \le 23$~kpc. 
For the gas  and the  stellar surface density distribution we consider the azimuthal averages  
shown in Figure ~\ref{surface}. Given the
30$\%$ uncertainty for the stellar mass surface densities  we shall
consider total M$_*$ in the following intervals: $9.57 \le$ log M$_* \le 9.81$~M$_\odot$ when using the $BVIgi$ 
mass map and $9.62 \le$ log M$_* \le 9.86$~M$_\odot$ when using the BVI   mass map. 
The vertically uniform gaseous disk is assumed with half thickness of 0.5~kpc and a flaring disk is considered for the
stellar disk with a half thickness of 100~pc at the center, reaching 1~kpc at the outer disk edge.
The contribution of the disk mass components to the rotation curve is computed  according to
\citet{1983MNRAS.203..735C}, who generalizes the formula for the radial force to thick disks. 
We use the reduced chi-square statistic, $\chi^2$, to judge the goodness of a model fit.

\subsection{Collisionless dark matter: a comparison with LCDM simulations}

The NFW density profile is usually written as: 

\begin{equation} 
\rho(R)={\rho_{NFW} \over {R\over R_{NFW}}\Bigl(1+{R\over R_{NFW}}\Bigr)^2} 
\end{equation} 

\noindent 
where $\rho_{NFW}$ and $R_{NFW}$ are the characteristic density and scale radius, 
respectively. Numerical simulations of galaxy formation find a  
correlation between $\rho_{NFW}$ and $R_{NFW}$ which depends on the  
cosmological model \citep[e.g. ][]{1997ApJ...490..493N,2001ApJ...559..516A, 
2001ApJ...554..114E,2001MNRAS.321..559B}. 
Often this correlation is expressed using the concentration parameter  
$C\equiv R_{vir}/R_{NFW}$ and $M_{h}$ or $V_{h}$.  
$R_{vir}$ is the radius of a sphere containing a mean density $\Delta$  
times the cosmological critical density. $\Delta$ varies between 93 and 97 depending
upon the adopted cosmology \citep{2008MNRAS.391.1940M}. This corresponds to a
mean halo overdensity of about 360  with respect to the cosmic matter density at z=0. 
$V_{h}$ and $M_{h}$ are the  characteristic velocity and mass at $R_{vir}$. In this paper 
we compare our best fitted parameters C and M$_{h}$ with the results of 
N-body simulations in a flat $\Lambda$CDM cosmology using relaxed halos for  
WMAP5 cosmological parameters  \citep{2008MNRAS.391.1940M}. 
In particular, we shall refer to
the relation between the mean concentration and virial mass of dark halos 
resulting  from the numerical simulation of \citet{2008MNRAS.391.1940M} whose dispersion is  
$\pm 0.11$ around the mean of log$C$. The resulting
C--M$_{h}$ relation is similar to that found by \citet{2007MNRAS.378...55M} and more recently   
by \citet{2012MNRAS.423.3018P}.   
 
\begin{figure} 
\includegraphics[width=\columnwidth]{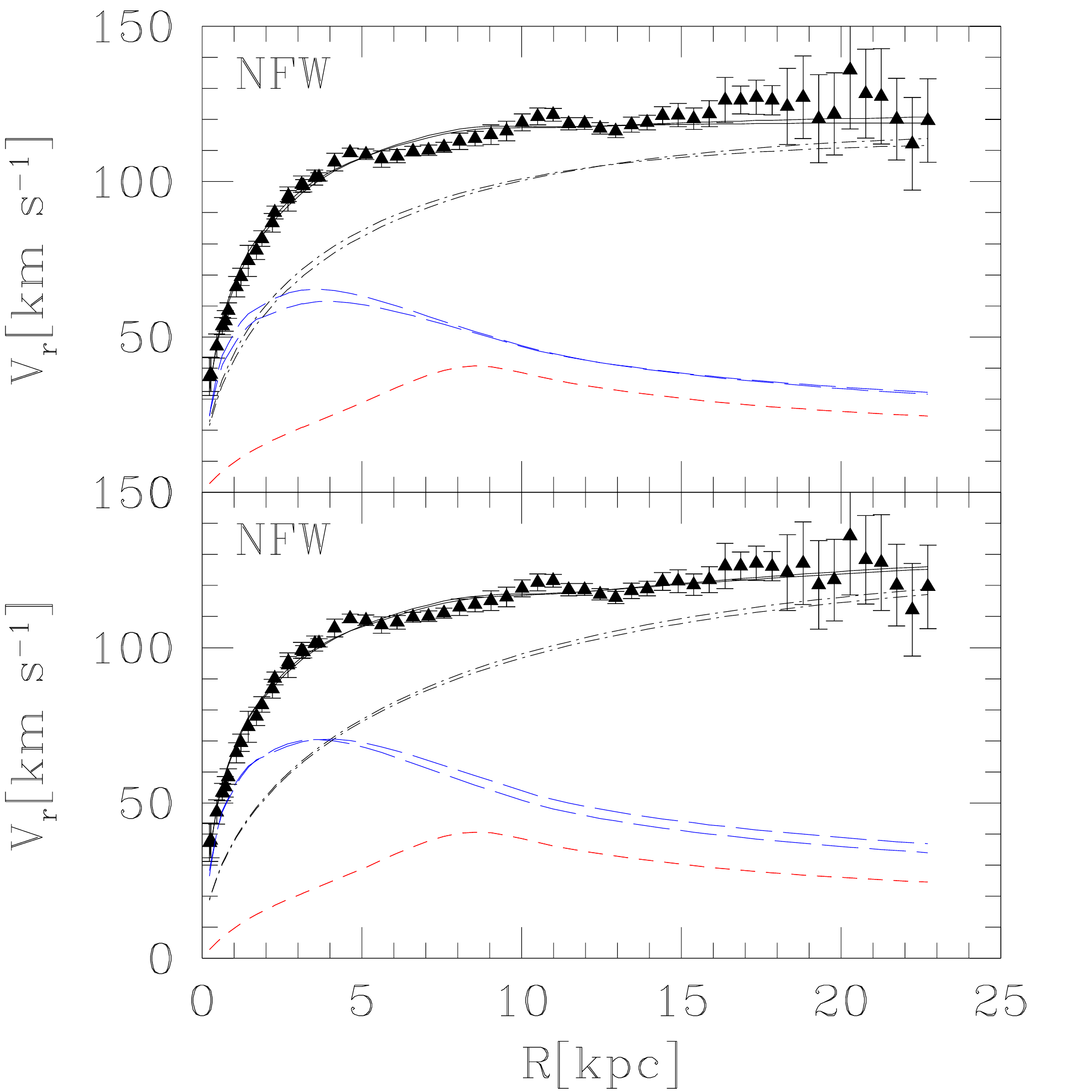} 
\caption{The rotation curve of M33 (filled triangles) and  best fitting dynamical models (solid line)
for NFW dark matter halo profiles (dot-dashed line). The small and large dashed lines (red and blue in the
on-line version) show respectively the gas and stellar disk contributions to the rotation curve. In the
{\it bottom panel} the highest stellar contribution and lowest dark halo curve are for the {\it best fitting 
dynamical model} using the $BVIgi$ mass map, the other curves are for the BVI mass map. The {\it top panel}
shows the most likely dynamical models for the two mass maps  when the likelihood
of the dynamical fit is {\it combined} with that of the stellar mass surface density and of the C--M$_h$ relation 
resulting from simulations of structure formation in a $\Lambda$CDM cosmology.}
\label{cdmfit} 
\end{figure}

\begin{figure} 
\includegraphics[width=\columnwidth]{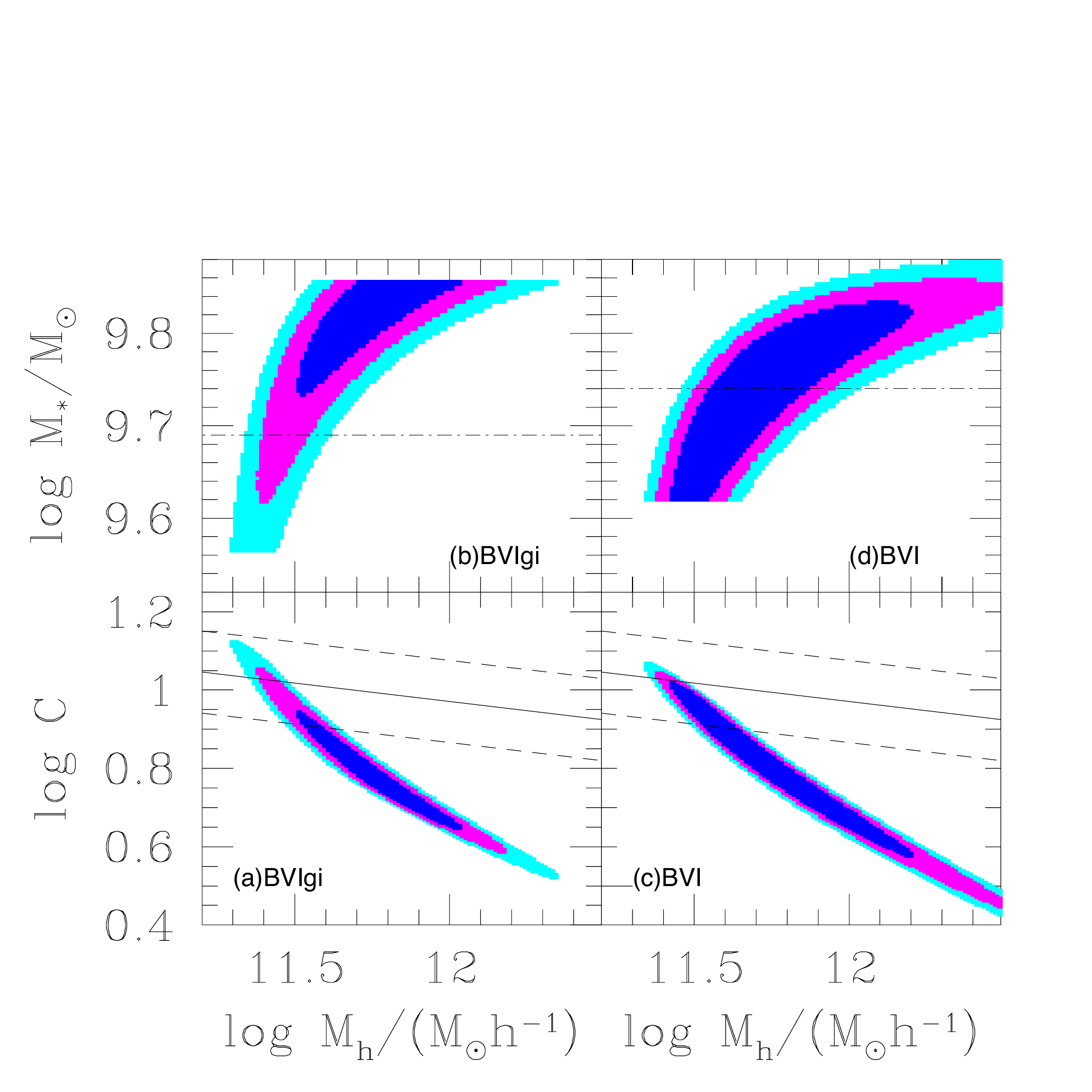} 
\caption{The 68.3$\%$ (darker regions, blue in the on-line version), 95.4$\%$ (magenta) and 99.7$\%$ (lighter regions, 
cyan) confidence areas in the log C--log M$_h/h^{-1}$ plane and in the M$_*$--M$_h/h^{-1}$ plane for the
{\it dynamical fit} to the rotation curve using a NFW dark matter halo profile. The mass distribution is that given  
by the $BVIgi$ colors for the areas shown in $(a)$ and $(b)$, and by the BVI colors for the areas in $(c)$ and $(d)$.
In panels $(a)$ and $(c)$ we show the C--M$_h$ relation (solid line) and its dispersion (dashed line) for $\Lambda$CDM
cosmology. In panel $(b)$ and $(d)$ the dash-dot line indicates the most likely stellar mass according the synthesis models which
rely on the colors following the panel labels. }
\label{cdmprob} 
\end{figure}

We now fit the M33 rotation curve using as free parameters the dark halo concentration, $C$, and 
mass, M$_{h}$. We fix the stellar mass surface density distribution to that given by our mass maps
extrapolated outwards, as explained in the previous section.   
We allow a scaling factor  to account for the model and data uncertainties i.e. we consider total stellar masses 
log~M$_*/M_\odot=9.69\pm 0.12$ and log~M$_*/M_\odot=9.74\pm 0.12$ for the  $BVIgi$ and  BVI mass distributions, respectively.
The best fits using the two mass maps are very similar as shown in the bottom panel of Figure~\ref{cdmfit}. 
The reduced $\chi^2$ are 0.96 and 1.08 for the BVI and $BVIgi$ stellar mass distributions respectively
with log~M$_*/M_\odot$=9.8, close to the upper limit of our considered range.
The dark matter halo for the best fits has concentration C=6.7 and mass M$_h$=5$\times 10^{11}$~h$^{-1}$~M$_\odot$. 
A close inspection of the 1,2,3-$\sigma$ confidence areas in the log C --log M$_h/h^{-1}$ plane and
in the log~M$_*$--log~M$_h/h^{-1}$ plane, shown in Figure~\ref{cdmprob}, reveals that indeed there is
still some degeneracy in the C--M$_h$ plane:  a lighter stellar disk  can provide good fits to
the rotation curve if a less massive dark halo with a higher concentration is in place. Confidence areas are 
traced in Figure~\ref{cdmprob} only for the allowed stellar mass range. In
Figure~\ref{cdmprob} we  also show the most likely value of the stellar mass computed by the synthesis 
models (dot-dashed line in the upper panels) and  the log C--log M$_h/h^{-1}$ relation as from 
$\Lambda$CDM numerical simulations \citep[continuous line][]{2008MNRAS.391.1940M}
and its dispersion around the mean (dashed lines). The values of C and M$_h$ suggested by the  
dynamical analysis of the  M33 rotation curve are in good agreement with the $\Lambda$CDM predictions.

\begin{figure} 
\includegraphics[width=\columnwidth]{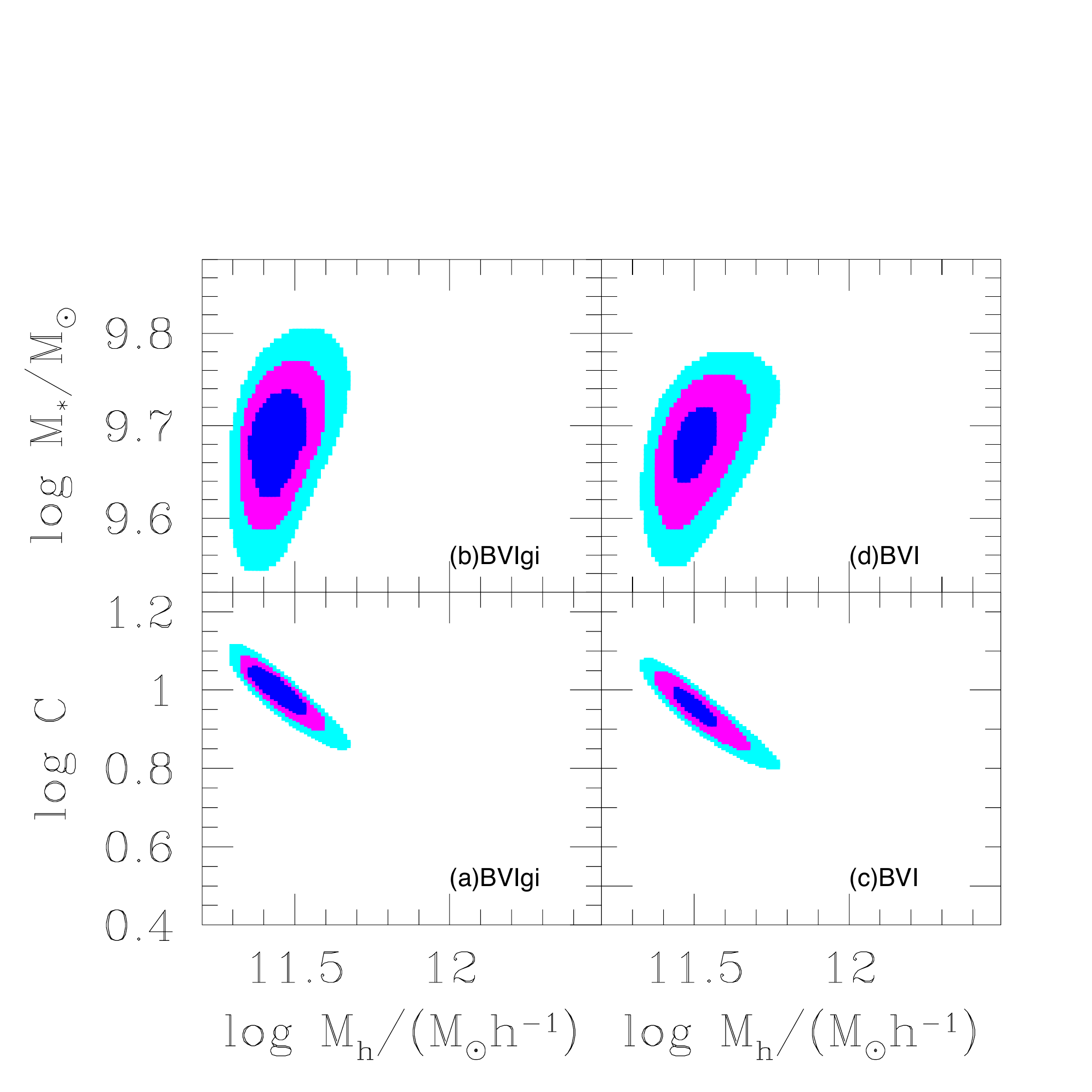} 
\caption{The 68.3$\%$ (darker regions, blue in the on-line version), 95.4$\%$ (magenta) and 99.7$\%$ (lighter regions, 
cyan) confidence areas in the log C--log M$_h/h^{-1}$ plane and in the M$_*$--M$_h/h^{-1}$ plane when the probability
of the dynamical fit is {\it combined} with the probability of the stellar mass surface density and  of the 
C--M$_h$ distribution from simulations of structure formation in a $\Lambda$CDM cosmology.The mass distribution is that given  
by the $BVIgi$ map for the areas shown in $(a)$ and $(b)$, and by the BVI map for the areas in $(c)$ and $(d)$.}
\label{probt} 
\end{figure}

We now compute the most likely values of the free parameters, C, M$_h$, and M$_*$ by considering the
composite probability of 3 events: $i)$ the dynamical fit to the rotation curve, $ii)$ the stellar mass determined
via synthesis models, $iii)$ the log C--log M$_h/h^{-1}$ relation found by numerical simulations
of structure formation in a $\Lambda$CDM cosmology.
The composite probability gives smaller confidence areas in the free parameter space, shown in  
Figure~\ref{probt}, which are very similar for the two mass maps. The model with the highest probability 
has the following parameters for the $BVIgi$ mass map and $h=0.72$:
M$_*=4.9^{+0.5}_{-0.7}\times 10^9~M_\odot$,  M$_h=3.9^{+1.0}_{-0.6}\times 10^{11}~M_\odot$,  C=$10\pm 1$, 
and the fit to the rotation curve, shown in the upper panel of Figure~\ref{cdmfit}, has a $\chi^2=1.3$.
For the BVI map we get  M$_*=4.8^{+0.4}_{-0.4}\times 10^9~M_\odot$,  
M$_h=4.6^{+0.7}_{-0.6}\times 10^{11}~M_\odot$,  C=$9\pm 1$ with a $\chi^2=1.1$ for the dynamical fit 
shown in the upper panel of Figure~\ref{cdmfit}.  Given the marginal differences in the two sets of free parameters
we can summarize the best fit $\Lambda$CDM dark halo model for M33
as follows:
 
\begin{equation}
M_h=(4.3{\pm 1.0}) \times 10^{11}~M_\odot \qquad  C=9.5\pm 1.5   
\end{equation}

\noindent
and the total stellar disk mass estimate is M$_*= (4.8 {\pm 0.6})\times 10^9$~M$_\odot$.
The resulting dynamical model implies that the contribution of the dark matter density
to the  gravitational potential is never negligible, although it becomes dominant outside 
the star forming disk (R$>7$~kpc) where the
stellar and gaseous disk gives a small, but similar contribution to the rotation curve.

In Table 1 (full version available in the on-line data) we display the rotation curve data, V$_{r}$,  
together with the azimuthal averages of the HI surface mass density, $\Sigma_{HI}$, and of the modelled surface mass density 
of the stars, $\Sigma_{*}$. 
The values of $\Sigma_{*}$ given in Table~1 correspond to  the most likely value of the stellar mass distribution according 
to BVIgi maps and to rotation curve fit for $\Lambda$CDM halo models: M$_*=4.8\times 10^9$~M$_\odot$.
 
\begin{table}
\caption{The rotation curve, the HI and stellar mass surface densities of M33}
\label{rotcur}
\begin{tabular}{lccccc }
\hline \hline
    R&  $\Sigma_{HI}$& $\Sigma_{*}$& V$_r$ & $\sigma_V$ \\
     (kpc)&  M$_\odot$~pc$^{-2}$ & M$_\odot$~pc$^{-2}$  & km~s$^{-1}$ &  km~s$^{-1}$ \\
 \hline \hline
       16.5 &  25.  & 25.  & 140.1 &  5.2 \\
       ... & ... & .... & ..... & .... \\
 \hline \hline
\end{tabular}
\end{table}

\subsection{Core models of dark matter halos}

The dark matter halo density profile proposed by \citet{1995ApJ...447L..25B} is a profile commonly used to
represent the family of cored density distributions \citep{2009MNRAS.397.1169D} and it is given by: 

\begin{equation} 
\rho(R)={\rho_B\over (1+{R\over R_B})\Bigl(1+({R\over R_B})^2\Bigr)} 
\end{equation}

\noindent 
where $\rho_B$ is the dark matter density of the core which extends
out to $R_B$ (core radius). The baryonic contribution to the M33
curve is declining  between 5 and 10~kpc. Hence, to have an independent estimate
of the dark matter density distribution according to \citet{2010A&A...523A..83S} we must  probe regions 
which are beyond 10~kpc. The extended rotation curve of M33, which increases by about 10-20$\%$ between 5~kpc 
and the outermost probed radius, is therefore appropriate for this purpose.
We have searched the best fitting parameters to the rotation curve, 
$\rho_B$, R$_B$, and M$_*$ (the last one within the limited range allowed by our stellar surface density maps).
There are no acceptable fits if the $BVIgi$ mass map is used, since the value of the minimum $\chi^2$ is 3.2.
We find  a minimum $\chi^2$ of 1.34 for the BVI mass map with  M$_*=7.2\times 10^9$~M$_\odot$ and the following
dark halo parameters:
 
\begin{equation}
R_B=7.5~kpc \qquad \rho_B=0.018~M_\odot~pc^{-3}  
\end{equation}

\begin{figure} 
\includegraphics[width=\columnwidth]{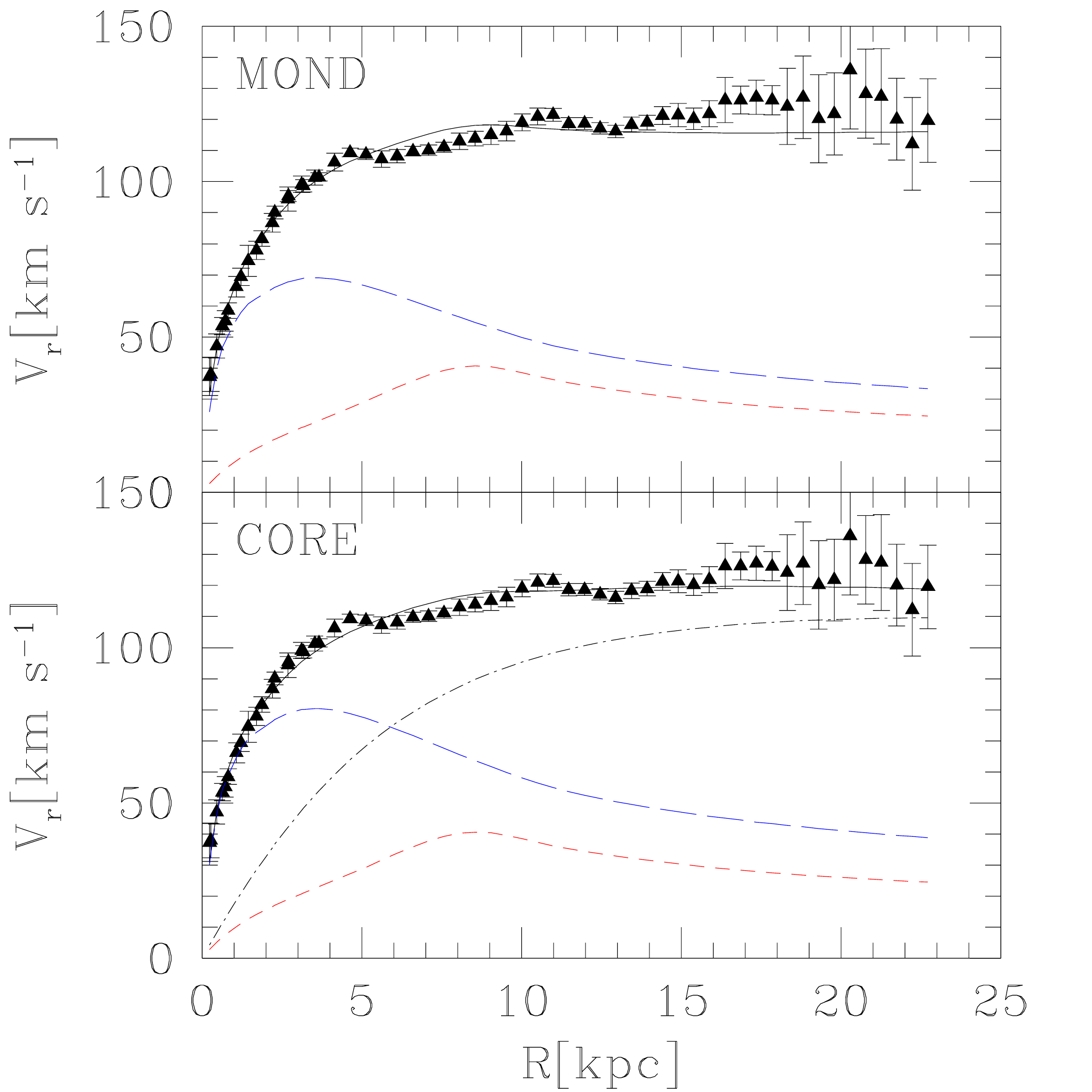} 
\caption{The rotation curve of M33 (filled triangles) and the best fitting model (solid line)
for a dark matter halo with a constant density core (bottom panel) and according to MOND (top panel). 
The small and large dashed lines (red and blue in the
on-line version) show respectively the gas and stellar disk Newtonian contributions to the rotation curves;
the dark halo contribution in the bottom panel is shown with a long dashed line.}
\label{coremond} 
\end{figure}

\begin{figure} 
\includegraphics[width=\columnwidth]{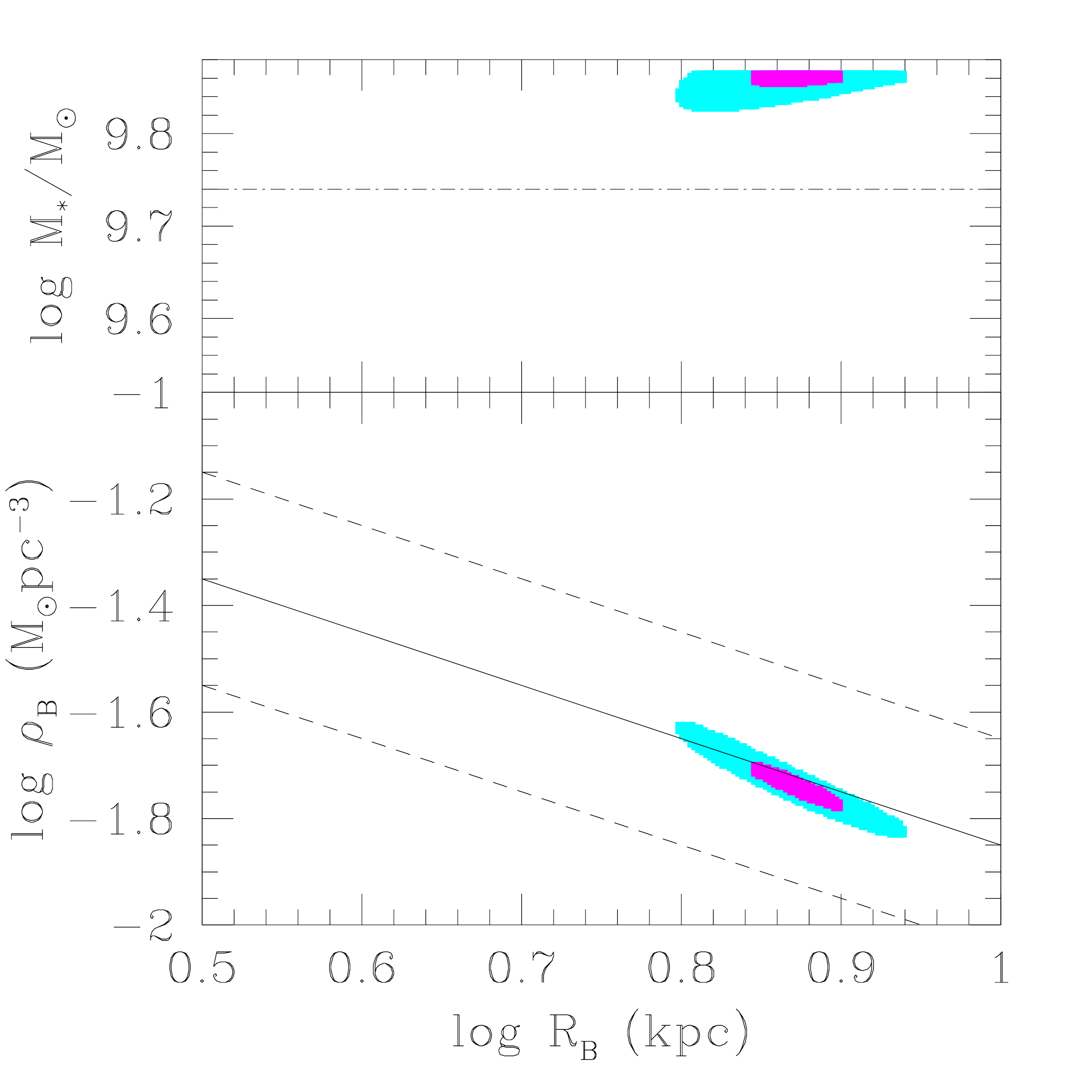} 
\caption{The  95.4$\%$ (magenta in the on-line version) and 99.7$\%$ (lighter regions, 
cyan) confidence areas in the log r$_0$ --log $\rho_0$ plane and in the log r$_0$ -- log M$_*$  plane relative to the
rotation curve fit using a core model and the BVI mass map. In the top panel
the dash-dot line indicates the most likely stellar mass according the synthesis models which
rely on the BVI colors. There are no core models in the 68.3$\%$ confidence area
for the BVI mass map and no acceptable core models for the $BVIgi$ mass map down to the 99.7$\%$ level. 
The solid line in the bottom panel
indicate the inverse log$\rho_0$ -- log r$_0$ correlation found by \citet{2009MNRAS.397.1169D}. }
\label{probcore} 
\end{figure}

\noindent
The best fit is shown in the bottom panel of Figure~\ref{coremond}. Let us notice
that in this galaxy we are able to probe the dark matter out to 3 times R$_B$.
Figure~\ref{probcore} shows the parameters in the 95.4$\%$  and 99.7$\%$ interval. Noticeably, the core
model provides acceptable fits when the  stellar mass is at the upper boundary of the interval 
compatible with the stellar surface density  map. A good quality fit for the $BVIgi$ stellar mass map 
with a halo core model requires lower core densities and about 10$^{10}$~M$_\odot$ of stars, which is extreme for
a galaxy like M33. This massive stellar disk implies a factor 2 higher stellar surface density than that computed here via 
synthesis models, unless the IMF has a Salpeter slope over the entire stellar mass range.  
However, despite the heavy stellar disk predicted by the cored dark matter density distribution, the
central surface density of the core is compatible with the value  
found by \citet{2009MNRAS.397.1169D}, log($\rho_B$ R$_B$/(M$_\odot$~pc$^{2}$))=2.15$\pm0.2$,
by fitting a very large sample of rotation curves of any luminosity
and Hubble type. We plan to investigate the consistency of a cored
dark matter halo with the M33 baryonic distribution in more detail in
a subsequent, dedicated paper.    
 
\subsection{Non-Newtonian dynamics without dark matter}

An alternative explanation for the mass discrepancy has been proposed by Milgrom   
by means of the modified Newtonian dynamics or MOND \citep{1983ApJ...270..365M}.  
Outside the bulk of the mass distribution, MOND predicts a much slower decrease of   
the (effective) gravitational potential, with respect to the Newtonian case. This    
is often sufficient to explain the observed non-keplerian behavior of RCs 
\citep{2002ARA&A..40..263S}. According to this theory, the dynamics becomes   
non-Newtonian below a limiting  acceleration value, ${a_0}$,  
where the effective gravitational acceleration takes the value  
$g = g_n/\mu(x)$, with ${g_n}$ the acceleration in Newtonian dynamics,
$x=g/a_0$, and $\mu(x)$ is an interpolating function between the Newtonian regime and
the case $g<<a_0$. Here, we shall use the critical acceleration value $a_0$    
derived from the analysis of a sample of rotation curves   
$a_0 = 1.2\pm 0.3\times 10^{-8}$~cm~s$^{-2}$ \citep{2002ARA&A..40..263S,
2011A&A...527A..76G}.
We have tested MOND for two choices of the interpolating function $\mu(x)$
\citep[see ][ for details]{2005MNRAS.363..603F}. In particular, we have used the
`standard' and the `simple' interpolation function and found that the former
provides better fits to the M33 data. 
Using the `standard' interpolating function, $\mu(x)=x/\sqrt{1+x^2}$, and the $BVIgi$
stellar surface density map, we find a minimum
$\chi^2=1.77$ just outside the 3-$\sigma$ confidence limits. For the best fit MOND requires
M$_*= 6.5 \times 10^9$~M$_\odot$ and  a$_0=1.4 \times 10^{-8}$~cm~s$^{-2}$.
A slightly lower $\chi^2$ (1.74) is found  using the BVI mass map with M$_*= 5 \times 10^9~M_\odot$
and a$_0=1.5\times 10^{-8}$~cm~s$^{-2}$. The fit provided by MOND does not improve considerably 
by increasing the stellar mass or the value of the critical acceleration a$_0$.

\subsection{The baryonic fraction}

The rotation curve of M33 is well fitted by a dark matter halo with a NFW density profile
and a total mass of (4.4$\pm$ 1.0)~10$^{11}$~M$_\odot$. This is much larger than the
halo mass which in M33 gives a baryonic fraction  equal to the cosmic value, f$_c=0.16$  
\citep{2003ApJS..148..175S}.
Taking the best fitting value of stellar mass we compute a baryonic fraction of about 0.022.
A baryonic fraction lower than the cosmic value is of no surprise since this is a common
results in low luminosity galaxies. Feedback from star formation, such as supernovae  driven outflows, is
likely responsible for  such a low baryonic fraction. Intergalactic filaments are in fact
enriched of metals thanks to the exchange of matter between galaxies and their environment.
Since the loss of gas from the galaxy will depend on the depth of  the potential well,
we expect that low-mass halos will be more devoid  of baryons.
The relationship between the stellar mass  of galaxies 
and the mass  of the dark matter halos has been derived by a statistical approach  matching
N-body simulated halo abundances as a function of the  mass to the observed abundance of galaxies 
as a function of their stellar mass \citep{2010ApJ...710..903M}. The resulting stellar mass fraction 
is  an increasing function of the halo mass up to  10$^{12}$~M$_\odot$. 
The analysis of the rotation curves is a different way of testing this scenario. The dark halo 
and stellar masses resulting from the dynamical analysis of the M33 rotation curve are compatible with the
statistical relationship found by \citet[][see their Figure~6]{2010ApJ...710..903M}  when scatter is taken into account.
A related question is whether the presence of outflows in the early evolutionary phases of M33
might have  affected the dark matter NFW profile. The recent work by \citet{2014MNRAS.437..415D,2014MNRAS.441.2986D}
has shown that the M33 halo mass is just at the edge of where its inner density profiles is expected to be modified  
by baryonic feedback, with a cuspy-like preferred inner slope.  
 
\section{Summary and conclusions}

The advantage of studying a nearby galaxy such as M33 is the possibility of  combining high resolution 
21-cm datasets with a overwhelming amount of multifrequency data. In  this work we
took advantage of existing  wide-field optical images in various bands to construct a map of 
stellar mass surface density. Two different sets of images, namely
the Local Group Survey \citep{2006AJ....131.2478M} and the SDSS \citep{SDSS}, have been combined
in the innermost 5~kpc. Further out, these optical images are not sensitive enough to constrain the
stellar surface density of M33 against variations in the background light. Thus we make use of the Spitzer 3.6~$\mu$m
map and of deeper observations with large optical telescopes \citep[e.g.][]{2011A&A...533A..91G} to 
estimate the stellar surface density scalelength out to the edge of the HI map. Using several 
methods for estimating the radially varying spatial orientation of the M33 disk, we derive the radial 
surface density distribution of the atomic gas and the rotation curve. By extrapolating the orientation 
of the outermost fitted ring for a few kpc outwards, we trace the rotation curve out to R=23~kpc.
The stellar and atomic gas maps,
together with the available informations on the molecular gas distribution, have been used to
derive  the baryonic surface density perpendicular to the galactic plane with small uncertainties. 
The knowledge of the potential well due to the baryons constrains the dynamical analysis 
of the rotation curve and the dark matter halo models more tightly
than previously possible. The radial distribution
of the stellar mass surface density inferred from the maps and comparison with the light distribution
emphasizes the importance of combining the dynamical analysis with
synthesis models.  In fact, the mass-to-light
ratio has non-negligible radial variations in the mapped region; additionally, local variations
are present in the disk such as between the arm and interarm regions.

Numerical simulations of hierarchical growth of structure in a $\Lambda$CDM cosmological model give 
detailed predictions of the dark matter density distribution inside the halos. The universal NFW radial profile
provides an excellent fit to the M33 rotation curve.  The free parameters, halo concentration and halo mass, 
are found to be C=9.5$\pm 1.5$ and M$_{h}$=4.3$\pm 1.0 \times 10^{11}$~M$_\odot$, when  
the  C--M$_{h}$ relation resulting from numerical simulations 
\citep{2008MNRAS.391.1940M} and the stellar mass surface density distribution via synthesis models are
taken into account. The best estimate of the stellar mass of M33 is   M$_*= 4.8 \times 10^9$~M$_\odot$,
with 12$\%$ residing in the outer disk. 
When added to the gas  this gives a  baryonic fraction 
of order of 0.02. A comparison of this  baryonic fraction with  the cosmic inferred value suggests an  
evolutionary history which should account for a loss of a large fraction of the original baryonic content. 

\begin{figure} 
\includegraphics[width=\columnwidth]{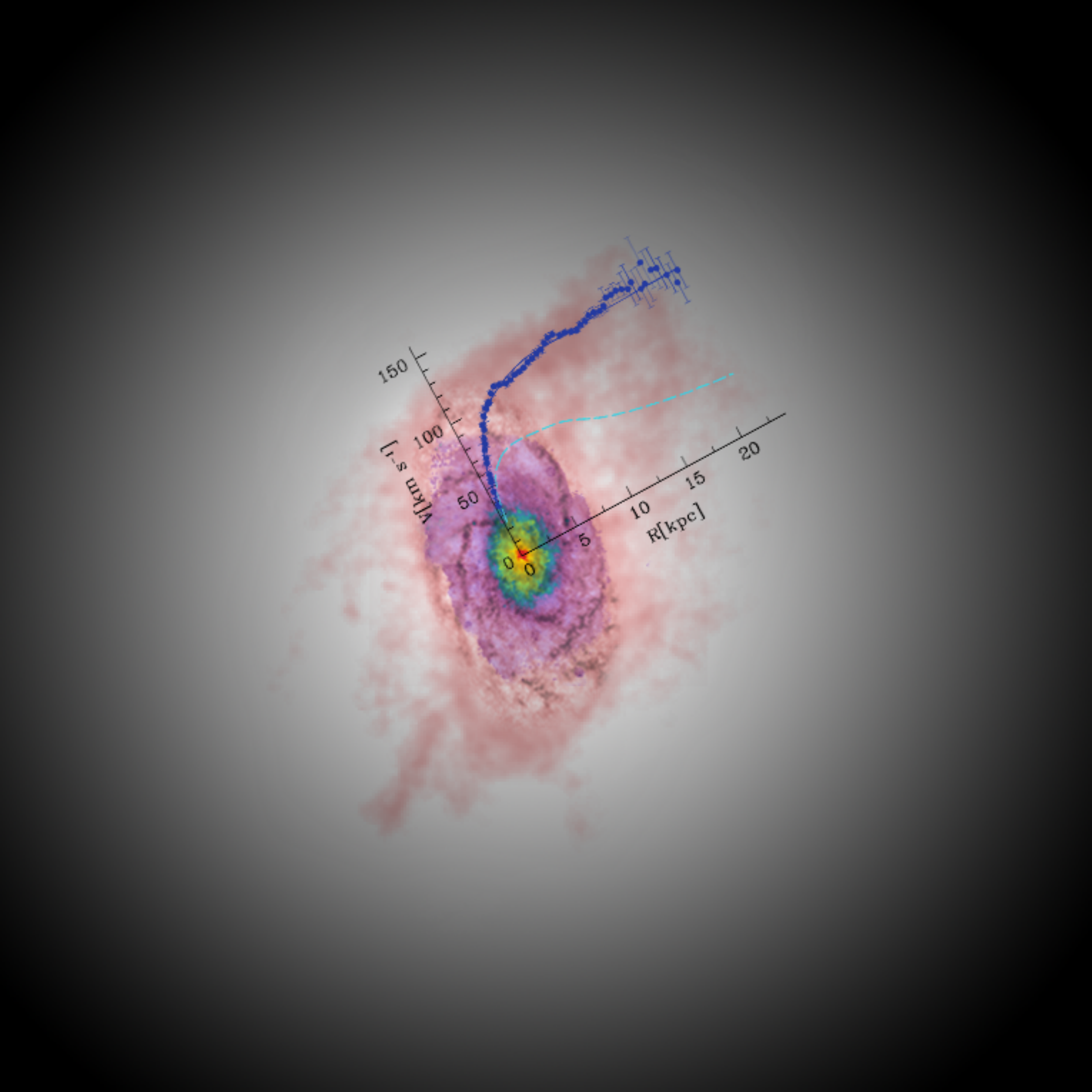} 
\caption{A  naive  view of M33: the inner stellar surface density (from red to magenta colors in the 
on-line version) superimposed to the HI gas distribution in log-scale (light pink ) are shown together with the 
stellar+gas contribution to the rotation curve (dashed cyan line) and with observed rotation curve (filled dots,
in blue). The analysis presented in this paper has pinned down the average dark matter density distribution
in the halo (light gray) which explains the observed rotation curve. The figure extrapolates the halo beyond  
the region analyzed in this paper in a naive smooth fashion.   
The total mass model  fit to the rotational data,  through the region analyzed in this paper, 
is shown by the continuous blue line. }
\label{picture} 
\end{figure}

A naive  view of the distribution of the baryons inside the NFW dark matter halo of M33,   as modelled
 in this paper, and the rotation curve are shown in Figure~\ref{picture}. 

The baryonic matter distribution in the framework 
of the modified Newtonian dynamic (MOND) does not provide good fits to the M33 data once the stellar content
is constrained. A dark halo with a constant density core is marginally compatible with the stellar mass distribution
and with the dynamical analysis of the M33 rotation curve, but requires a heavy stellar disk at the limit of the
range allowed by our mass maps.  The presence of a dark cuspy core in M33, as predicted by structure
formation in a $\Lambda$CDM hierarchical universe, is in agreement with  numerical simulation
of the baryonic feedback effects on the density profiles of dark matter haloes \citep{2014MNRAS.437..415D}. 
For our best fit halo mass, energy from stellar feedback is insufficient 
to significantly alter the inner dark matter density, and the galaxy retains a cuspy profile. Given the uncertainty
in the M33 stellar-to-dark matter ratio, however, we cannot exclude  a slight modification of the original profile.

Having determined the stellar mass of M33 and considering a star formation rate of 0.45~M$_\odot$~yr$^{-1}$ 
\citep{2009A&A...493..453V} at z=0, we estimate a specific star formation rate (SSFR) of about 10$^{-10}$~yr$^{-1}$,
in  agreement with the value inferred 
for large galaxy samples at z=0 and by  multi-epoch abundance matching models \citep[e.g.][ ]{2013MNRAS.428.3121M}.
A future analysis of the map of the SSFR of M33, in the framework of a chemical evolution model which
takes into account gas outflow and inflow (according to the inferred dark halo mass) and reproduces the observed metallicity 
gradients \citep{2010A&A...512A..63M}, will provide useful insights on the star formation history  and, more 
in general, on galaxy evolutionary models.    

\begin{acknowledgements}
We are grateful to St\'ephane Charlot and Gustavo Bruzual for kindly providing us with the latest revision of the BC03 stellar 
population synthesis models ahead of publication. We aknowledge financial support from PRIN MIUR 2010-2011, project 
``The Chemical and Dynamical Evolution of the Milky Way and Local Group Galaxies'', prot. 2010LY5N2T.
\end{acknowledgements} 

\begin{appendix}
\section{The tilted-ring model}
\label{appa}

To better constrain the disk orientation using the VLA+GBT 21-cm datacube we first smooth spatially the data 
described in the  previous section at 130~arcsec resolution  in order to gain sensitivity. 
At this spatial resolution, we reach a brightness sensitivity of 0.25~K. Considering a typical signal width of 
20~km~s$^{-1}$ our sensitivity should be appropriate for detecting HI  gas at column densities as low as  10$^{19}$~cm$^{-2}$. 
To determine the disk orientation, we compare the tilted
ring models directly against the full spectral database considering channels 
12.88~km~s$^{-1}$ wide (rather than fitting the moments of the flux distribution). The cube consists of  2475 positions 
(i.e. pixels $40\times40$~arcsec$^2$ wide) in which 21-cm emission has been detected, and for each position we have
25 velocity channels covering from -20 to -342 km~s$^{-1}$ heliocentric velocities.  
We summarize below the main features of our method. 

We use 110 tilted concentric rings in circular rotation around 
the center to represent the overall distribution of HI. Each ring is characterized by its radius $R$ and by 7 
additional parameters: the H I surface density $S_{HI}$, the circular velocity $V_c$, the inclination 
$i$ and the position angle PA with respect to the line of sight, the systemic velocity $V_{sys}$ and 
the position shifts of the orbital centers with respect to the galaxy center 
($\Delta x_c, \Delta y_c$). These last 3 parameters allow the rings to 
be non concentric and to have velocity shifts with respect to the systemic due to local perturbations 
(such as  gas outflowing or infalling into the ring or M31 tidal pull). Of these large set of rings 
we allow only the parameters of 11 equally spaced rings, called the "free" rings, to vary independently. 
We set the properties of the 1st  
innermost free ring to be the same as those of the 2nd  ring (because due to its small size
it turns out to be highly unconstrained) and we keep the free rings to be the 11,22,33...110th ring. 
The properties of rings between each of the free ring radii were then linearly interpolated. Each of the 7 
parameters of the 10 free rings were allowed to vary.  
We assume that the emission is characterized by
a Gaussian line of width $w$, which is an addition free parameter of the model,  centered at V$_c$. We compute 
the 21-cm emission along each ring as viewed from our line of sight, and the synthetic spectrum at each pixel by convolving
the emission from various ring pieces with
the beam pattern. We then test how well the synthetic and observed spectra match by comparing the flux densities in 
25 velocity channels. With this method we naturally account for the possibility that the line flux in a pixel 
might result from the superposition along the line of sight of emission from different rings.  
As initial guess for the free parameters of the tilted ring model  we follow the
results of  \citet{1997ApJ...479..244C}.  

The assignment of a measure of goodness of the fit is done  following two methods: the 'shape' and the 'v-mean' method.
In the shape method we minimize a $\chi^2$ given by the sum of two terms, the flux and the shape term. The flux term is set 
by the difference between the observed and modeled fluxes in each pixel. 
The shape term retains information about the line shape only, that are lost when just the first few moments are examined. 
This is essential in the regions of M33 where the emission is non-gaussian, for example when the velocity distribution 
of the gas is bimodal (this is indeed the case for some regions in the outer disk of M33, see CS). The shape term
is given by the difference between the observed and the normalized modeled fluxes in each pixel and  for every 
spectral channel. The normalized  model spectrum is the flux predicted by the  model in a given  channel multiplied by
the ratio of the  observed to model integrated emission. In doing so the shape
term is no longer dependent on the total flux. The shape term is computed only for pixels with flux larger than 
0.2 Jy~km~s$^{-1}$/beam i.e. N$_{HI}\ge 1.3\times 10^{19}$~cm$^{-2}$.
The error for the shape term is the rms in the baselined spectra, $\sigma_{n,i}$, which is the experimental uncertainty 
on the flux in the $i$-channel at the $n$-pixel. The noise in each channel of the datacube, $\sigma_{ch}$, is uniform  
and estimated to be 0.0117 Jy/beam/channel.  
As suggested by \citet{1997ApJ...479..244C} we  estimate $\sigma_{n,i}$ as:

\begin{equation}
\sigma_{n,i}= 2 {w_i\times \sigma_{ch}\over \sqrt{w_i/2.57}}
\end{equation}

\noindent
where $w_i=12.88$~km~s$^{-1}$ is the width of the channel used to compare the data with the model and 2.57 km~s$^{-1}$ is the 
database channel width.
The  flux term is affected by the uncertainty on the integrated flux and by calibration uncertainties, proportional 
to the flux. The calibration error forces the minimization to be sensitive to weak-line regions and is 5$\%$.
The resultant reduced  $\chi^2$  formula is the following: 

\begin{equation}
 \chi^2_{shape}  = {1\over {N-N_f}}\sum_{n=1}^N\bigg\lbrace{(I_n^{mod}-I_n^{obs})^2\over 25\sigma_{n,i}^2+(0.05 I_n^{obs})^2 } 
\end{equation}
\begin{equation}  
   + {1\over 25}\sum_{i=1}^{25} {\lbrack(I_n^{obs}/I_n^{mod})I_{n,i}^{mod}-I_{n,i}^{obs}\rbrack^2\over  \sigma_{n,i}^2 }
\bigg\rbrace \\
\end{equation}

where N is the number of pixels (2475)  and N$_f$ is the
number of free parameters (71 for our basic model), leaving a large number of degrees of freedom.  
Given the difficulty in finding a unique minimum  we use a two step method  to converge toward the minimal solution. 
Since some of the parameters might be correlated we begin by searching for minima over a grid of the parameters 
surrounding our first guess. 
We carried out several optimization attempts under a variety of initial conditions and with different orderings 
for adjusting the parameters. After iterating to smaller ranges of variation, we choose the parameter values which gives the 
minimum $\chi^2$. In the second step we check the minimal solution by applying a technique of partial minima. 
We evaluate the $\chi^2$ by varying each parameter separately.
We checked our solution by surrounding  the galaxy with  zero-flux observations for stabilizing the outermost ring. 
It is important to notice that the flux and the shape term give a similar contribution to the minimum $\chi^2$ value.

In the second method  we determine a solution using the deviations of the integrated flux and of the intensity-weighted mean 
velocity along the line of sight  at each pixel. We carried this out with another two-step procedure, allowing all 71 
parameters to be varied. We start by keeping  the ring centers and 
systemic velocity fixed; then  the rings centers and velocities are considered as free parameters in the minimization as well.
In order to keep the model sensitive to variations of parameters
of the outermost rings, each pixel in the map is assigned equal weight. Pixels with higher or lower 21-cm surface brightness 
contribute equally to determine the global goodness of the model fit. 
Since the original data  has a velocity resolution of 1.25~km~s$^{-1}$, we arbitrarily set
$\sigma_m$, the uncertainty in the mean velocity, equal to the width of about 5 channels ($\sim$6~km~s$^{-1}$). 
This is simply a scaling factor which gives similar weight to the two terms in the $\chi^2$ formula. 
The equation below defines the  reduced $\chi^2$ of the v-mean method:

\begin{equation}
\chi^2_{mean} = {1\over {N - N_f}} \sum_{n=1}^N\bigg\lbrace{(I_n^{mod}-I_n^{obs})^2\over 25\sigma_{n,i}^2+(0.05 I_n^{obs})^2 }  
\end{equation}
\begin{equation}
 +  (V^{mod}_n-V^{obs}_n)^2/\sigma_m^2  
\end{equation}
 
\noindent
where V$_{mod}$ is the mean velocity predicted by the tilted ring model at the  pixel $n$.

\begin{figure} 
\includegraphics[width=\columnwidth]{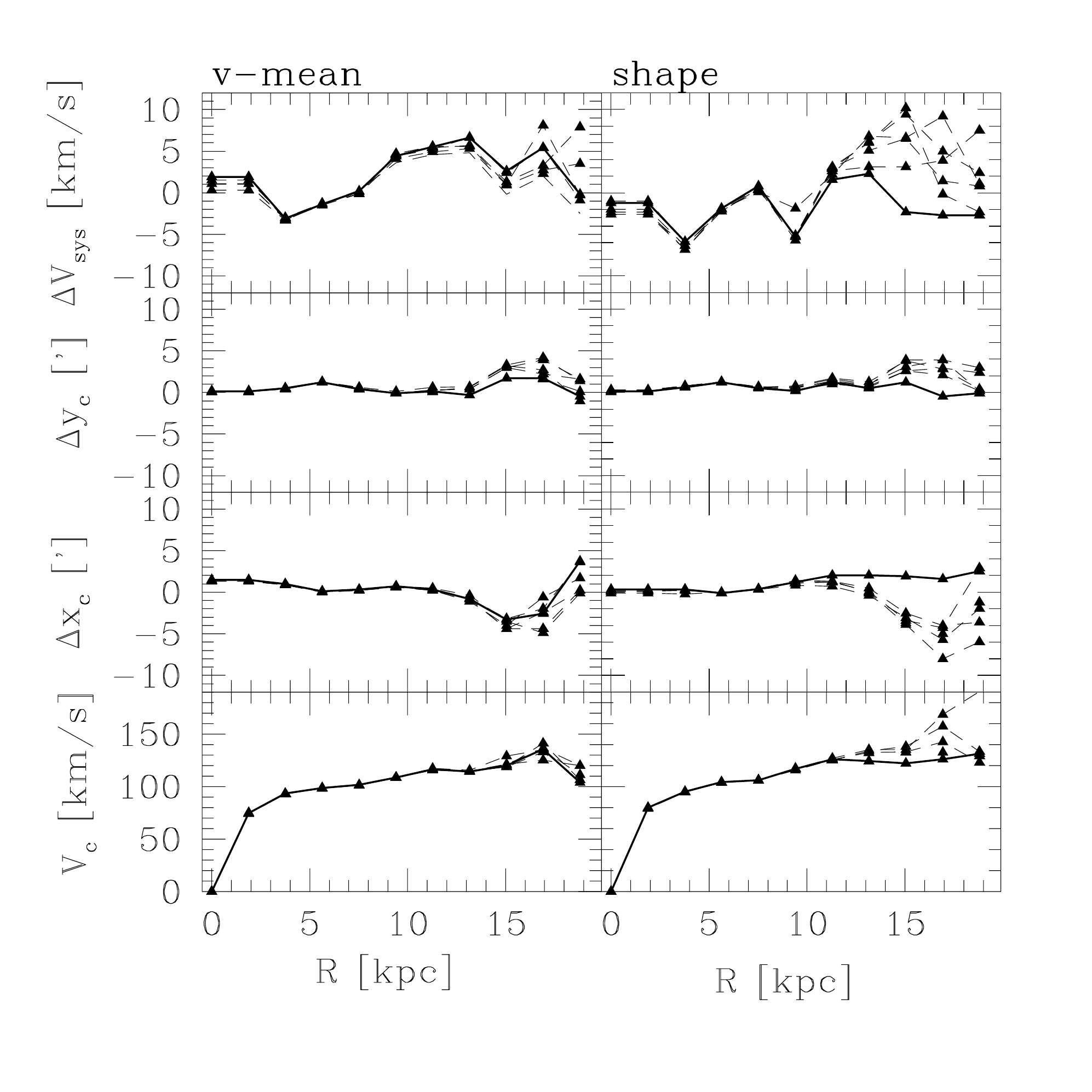} 
\caption{The parameters V$_{c}$, $\Delta$x$_c$, $\Delta$y$_c$ 
and $\Delta$V$_{sys}$ 
corresponding to an assortment of tilted ring models whose $\chi^2$ is within 2$\%$ of the absolute minimum value found.
In the panels to the left the parameters for the v-mean minimization are shown, to the right those for the shape
minimization. The continuous line connects the parameters of the best fit tilted ring models used
for deriving the rotation curve. }
\label{ringmodels} 
\end{figure}

Given the large number of degrees of freedom, the increase of $\chi^2$ corresponding to 1-$\sigma$  probability interval 
for Poisson statistics would be very small. The $\chi^2$ standard deviation  is of order 0.03 for the flux term and of order
0.006 for the shape term. Since the  presence of local perturbations does not allow the model to approach a $\chi^2$ of order unity,
we consider   $\chi^2$ fractional variations corresponding to the mean value of the two terms (i.e. 2$\%$). 
By testing the $\chi^2$ variations for each variable independently
we should have an indication on which ring and parameter is well constrained by the fitting procedure. 
Hence we first arbitrarily collect all possible sets of tilted ring models which give 
local minima in the $\chi^2$ distribution with values within 2$\%$ of the lowest $\chi^2$ (which is 7.3 and 6.8 for the v-mean and 
shape method respectively). In Figure~\ref{ringmodels} we show V$_{c}$, $\Delta$x$_c$, $\Delta$y$_c$ and $\Delta$V$_{sys}$ 
corresponding to an assortment of models whose $\chi^2$ is within 2$\%$ of the absolute minimum value found. 
The adopted systemic velocity is $V_{sys}$=-179.2 km/s. The displacements 
of the ring centers and systemic velocities are not very large and  increase going radially outwards, as does the scatter 
between solutions corresponding to partial minima.
The value of the velocity dispersion we find from the minimization is of 
order 10~km~s$^{-1}$. 

For each minimization method we then select a tilted ring model between those with acceptable $\chi^2$
using the maximum North-South symmetry criterion for  rotation curves relative to the two galaxy halves. The
corresponding values of $i$ and PA are shown in Figure~\ref{pa-i} of Section 4 with the relative uncertainties.
The uncertainties are computed by varying one parameter of each free rings at a time,  around the minimal solution 
until the $\chi^2$ increases by 2$\%$. Simultaneous parameter variations within the given 
uncertainties gives $\chi^2$ variations larger than 2$\%$.
In deriving the rotation curve  we take into account the  uncertainties considering deconvolution models in which 
the inclination or PA of all the rings vary simultaneously. In this case  we consider only 35$\%$ of the uncertainties displayed 
in Figure~\ref{pa-i}, in either PA or $i$, in order to have a $\chi^2$ within 2$\%$  of the minimal solution.

\section{Finite disk thickness corrections}
\label{appb}

The "rotation curve" is the azimuthal component
of the velocity in the equatorial plane of the disk at given 
galactocentric distance. However, the 21-cm spectrum observed at a certain position
in an inclined disk depends not only on the azimuthal velocity in the plane but also on
two additional effects: the smearing due to the extent of the telescope beam,
and the vertical extension of the disk; both become more severe with
increasing inclination.
It is worth noticing that the tilted ring model fit runs over
a smooothed database, spatially and in frequency, whose final geometrical parameters are
then used to derive the rotation curve from a higher resolution datasets. Therefore instead
of including disk finite thickness effects in the tilted ring models we prefer to 
account for this and for the beam smearing in the 21-cm spectral
cube at the original resolution  using a a set of numerical simulations. 
 
We assume the gas to be in an azimuthally symmetric disk inclined
with respect to the line of sight according to the tilted ring model fit. 
The gas radial distribution is set equal to that given
by the  integrated spectral profile, while the vertical one is 
modeled by an exponential with a folding length of 0.3~kpc. 
Only a  disk component is considered with no allowance for 
a halo component \citep{2007AJ....134.1019O}.
For the beam we used a gaussian with $FWHM=20$~arcsec truncated at a
radius of 24~arcsec; the channel spacing in the spectrum is 1.25~km~s$^{-1}$.
The input rotation curve is the one obtained by the observed peak
and mean-velocities, and the random velocity is assumed to be isotropic and spatially 
constant with $FWHM=8.0$~km~s$^{-1}$, as observed in most of the disk. 
We ran simulations with and without a vertical
rotation lag according to \citet{2007AJ....134.1019O}.

Using the above parameters, we simulate  
a synthetic HI data cube. The cube is  used to derive for each position
the corresponding velocity, either the peak or the flux-averaged mean,
and then build a simulated rotation curve. 
The differences between the simulated and the input rotation give the average 
corrections to the rotation curve as function of radius. We shall refer to these corrections
as {\it finite disk thickness} corrections. They are
of order 2-3~km~s$^{-1}$ and reach values of 5~km~s$^{-1}$  only within the
innermost 200~pc. We apply these corrections to the rotational velocities.

\end{appendix}

\end{document}